\documentclass[preprint]{elsarticle} 

\usepackage[utf8]{inputenc}
\usepackage{amsmath}
\usepackage{mathtools}
\usepackage{mathalfa}
\usepackage{mathrsfs}
\usepackage{amssymb}
\usepackage{dsfont}
\usepackage{graphicx}
\usepackage{float}
\usepackage{physics}
\usepackage{amsfonts}
\usepackage{amsthm}
\usepackage{siunitx}
\usepackage{microtype}
\usepackage{cancel}
\usepackage{bm}
\usepackage[hidelinks=true, hyperfootnotes=true, bookmarks=true]{hyperref}
\usepackage{mathalfa}
\usepackage{mathrsfs}
\usepackage{verbatim}
\usepackage{import}
\usepackage{multicol}
\usepackage{mdframed}
\usepackage{todonotes}
\usepackage{listings}
\usepackage{arydshln}
\usepackage{gensymb}

\biboptions{authoryear}

\usepackage[capitalise, nameinlink]{cleveref}

\usepackage{subfigure}
\usepackage{caption}

\hypersetup{colorlinks={true},linkcolor={red},citecolor=blue}

\newcommand{\dpart}[2]{\frac{\partial #1}{\partial #2}}

\renewcommand{\vec}[1]{\bm{#1}}

\newcommand{\parentesi}[1]{\left(#1\right)}
\newcommand{\claudator}[1]{\left[#1\right]}

\newcommand{\limitss}[2]{_{#1}^{#2}}

\usepackage{todonotes}
\usepackage{xcolor}
\usepackage{soul}

\definecolor{ao(english)}{rgb}{0.0, 0.5, 0.0}

\usepackage[a4paper, left=30mm, right=30mm, top=30mm, bottom=30mm]{geometry}

\journal{Ecological Modelling} 

\begin{document}

\begin{frontmatter}

\title{Modelling Parasite-produced Marine Diseases: The case of the Mass Mortality Event of \textit{Pinna nobilis}}


\author[1]{Àlex Giménez-Romero}

\author[2,3]{Amalia Grau}

\author[4]{Iris E. Hendriks}

\author[1]{Manuel A. Matias\corref{cor}}
\ead{manuel@ifisc.uib-csic.es}

\cortext[cor]{Corresponding author}

\address[1]{IFISC (CSIC-UIB), Instituto de Física Interdisciplinar y Sistemas Complejos, Campus UIB\\ E-07122 Palma de Mallorca, Spain}
\address[2]{Laboratori d’Investigacions Marines i Aqüicultura (LIMIA), Govern de les Illes Balears, Av. Gabriel Roca 69\\ E-07158 Port d’Andratx, Mallorca, Spain}
\address[3]{INAGEA (INIA-CAIB-UIB), 
Campus UIB, E-07122 Palma de Mallorca, Spain}
\address[4]{Instituto Mediterráneo de Estudios Avanzados, IMEDEA (CSIC-UIB), E-07190 Esporles, Mallorca, Spain}

\begin{abstract}
The state of the art of epidemic modelling in terrestrial ecosystems 
is the compartmental SIR model and its extensions from the now classical work of Kermack-Mackendrick. In contrast, epidemic modelling of marine ecosystems is a bit behind, and compartmental models have been introduced only recently. One of the reasons is that many epidemic processes in terrestrial ecosystems can be described through a contact process, while modelling marine epidemics is more subtle in many cases. Here we present a model describing disease outbreaks caused by parasites in bivalve populations. The SIRP model is a multicompartmental model with four compartments, three of which describe the different states of the host, susceptible (i.e. healthy), S, infected, I, and removed (dead), R, and one compartment for the parasite in the marine medium, P, written as a $4$-dimensional dynamical system. Even if this is the simplest model one can write
to describe this system, it is still too complicated for both direct analytical manipulation and direct comparison with experimental observations, as it depends on four parameters to be fitted. We show that it is possible to simplify the model, including a reduction to the standard SIR model if the parameters fulfil certain conditions. The model is validated with available data for the recent Mass Mortality Event of the noble pen shell \textit{Pinna nobilis}, a disease caused by the parasite \textit{Haplosporidium pinnae}, showing that the reduced SIR model is able to fit the data. So, we show that a model in which the species that suffers the epidemics (host) cannot move, and contagion occurs through parasites, can be reduced to the standard SIR model that represents epidemic transmission between mobile hosts. The fit indicates that the assumptions made to simplify the model are reasonable in practice, although it leads to an indeterminacy in three of the original parameters. This opens the possibility of performing direct experiments to be able to solve this question.
\end{abstract}









\begin{keyword}
Epidemics \sep Mathematical model \sep Compartmental model \sep Bivalve diseases \sep \textit{Pinna nobilis} \sep \textit{Haplosporidium pinnae}
\end{keyword}

\end{frontmatter}

\section{Introduction} \label{sec:Introduction}
    
   Marine organisms, like their terrestrial counterparts, can serve as hosts for a diversity of parasites and pathogens present in the ecosystem, which are directly responsible for disease outbreaks. Disease induced mortality affects not only the host population, but can cascade through the whole ecosystem, altering its structure and functioning \citep{Ward2004}. Furthermore, climate change can increase the spread range and impact of parasites and pathogens \citep{Burge2014}. In fact, marine infectious diseases are recently increasing due to climate change and other anthropogenic pressures, like pollution and overfishing \citep{Lafferty2004}. This, in turn, threatens many valuable ecologically habitats and can also result in substantial economic losses in e.g. aquaculture \citep{Lafferty2015}. Analysing the impact of these events at appropriate scales (spatial and temporal) and biological organisation levels (species, populations and communities) is crucial to accurately anticipate future changes in marine ecosystems and propose adapted management and conservation plans \citep{Pairaud2014}. Thus, there is a strong need to address 
   the mechanism of disease propagation in marine organisms.
    
    However, the state of the art of epidemiological studies in marine ecosystems lags behind that of terrestrial ecosystems \citep{Harvell2004}.
    Contact and vector-borne based infectious diseases of terrestrial vertebrates and their epidemiology are typically studied using variations of the classical formulation of Kermack and McKendrick \citep{McKendrick, McKendrick_2, McKendrick_3}, the SIR model. Among other things, this formalism allows to understand why epizootics spread and stop, as the propagation of a disease is a threshold phenomenon \citep{Anderson1991}, regulated by the now commonplace  $R_0$ dimensionless number. Within this framework the initiation of epidemic transmission occurs when an infected individual is in close contact with a susceptible host or through a transmission vector, as typically pathogens can only survive for a very limited time outside the host in an aerial environment.
    On the other hand, as air is typically a much harsher medium for pathogens than water, the sea is expected to host a large number of pathogens (viruses, bacteria and parasites) for a relatively long time. The longer life span of pathogens in a water medium,
    together with the increased buoyancy arising from the different physical properties of seawater and air, coupled to the existence of marine currents that can transmit pathogens for long distances away, 
    allows diseases to spread faster and reach further distances in marine environments compared to epidemics in terrestrial systems \citep{CANTRELL2020239}.
    As a result, the possible long-term transmission of parasites by currents in marine environments make them more prone to suffer from persistent zoonotics compared to terrestrial ecosystems, where for an epidemic outbreak to occur the presence of an initial infected host (or vector) is necessary within a susceptible population.
    Until quite recently, marine zoonotics were mostly studied using different
    models compared to terrestrial diseases and it was not even clear whether the same tools could be applied \citep{MCCALLUM_intro}. 
    The abundance of pathogens in marine ecosystems is one of the reasons why proliferation models, that do not focus on transmission and assume a widespread occurrence of the pathogen and a rapid transmission problem, have been most popular in the field \citep{Powell2015}.
    In fact, compartmental models are starting to be used only recently in the study of marine epizootics \citep{article_SIP}.
    
    An important subset of marine organisms are sessile, e.g. bivalves, which means that they can not move.
    In the case of sessile terrestrial organisms disease transmission occurs mostly through vectors, insects that transmit the pathogens causing the disease. Instead, in marine ecosystems disease transmission is most often waterborne, in particular in passive water filtering feeders, as is the case of bivalves. Recently, some compartmental models considering the pathogen population have been proposed to study particular bivalve epidemics \citep{BIDEGAIN_2016_2, article_SIP, BIDEGAIN_perkinsus}. In the present work we analyse a model that is aimed to describe disease transmission from an infected immobile host to a susceptible one of the same species through waterborne parasites, that are explicitly described. The model is closely related to the SIP model introduced in \cite{article_SIP}. In this first study we analyse in detail the properties of the mean-field version of the model, that aims to describe spatially homogeneous (i.e. well mixed) populations. The well mixed approximation will be valid whenever the mean distance among hosts is smaller than the mixing length of the parasites before they get inactivated or absorbed.
    The model is written such that waterborne transmission is the only mechanism by which one infected immobile host can infect a healthy one, and, thus, does not describe infection through direct contact.
    It is also assumed that the infected hosts, as invertebrates, do not have immune memory, and that the probability that an infected individual recovers is small and can be neglected. 
    Thus, the model is not adequate to study infection of highly aggregated molluscs (like some mussels) or other passive filters like corals, as for these hosts one should also include the possibility of infection through direct contact. A first very relevant question is whether the model, describing infection of immobile (sessile) hosts through waterborne parasites can be reduced to a simpler version in which the parasite compartment is not needed. 
     One exact and two approximate reductions are presented. We believe that the model can be most useful in the rapid characterisation of emergent marine epidemics if the right data from a well mixed system are available.
     
        A very timely case study of such emerging epidemics is the noble fan mussel (or pen shell) \textit{Pinna nobilis}. This fan mussel is the largest endemic bivalve in the Mediterranean Sea, and is under a serious extinction risk due to a Mass Mortality Event (MME) that has occurred throughout the whole Mediterranean basin very recently \citep{March, ZOTOU2020, VAZQUEZ2017}. Right before this MME, it was distributed across a wide type of habitats including coastal and paralic ecosystems at depths between 0.5 to $\SI{60}{m}$ \citep{butler1993ecology, PRADO2020105220}. In open coastal waters, the distribution of the species is mainly associated with seagrass meadows, typically of \textit{Posidonia oceanica}, which has been indicated as its optimal habitat \citep{Hendriks2011}.
    Its lifespan is up to 50 years in favourable conditions and its size can get up to $\SI{1.2}{m}$, placing it among the largest bivalves of the  world \citep{Cabanellas2019}. These fan mussels play a crucial ecological role in their habitat, as \textit{P. nobilis} individuals filter water, thus retaining a large amount of organic matter from suspended detritus, contributing to water clarity \citep{TRIGOS2014}. Furthermore, it is a habitat-forming species, because its shell provides a hard-surface within a soft bottom ecosystem, which can be colonised by different benthic species, augmenting biodiversity \citep{Cabanellas2019}. In addition, at very dense populations, the species can function as an ecosystem engineer, creating biogenic reefs \citep{katsanevakis2016transplantation}. 
 
    Despite \textit{P. nobilis} populations have greatly declined due to anthropogenic activities in the $20$th century \citep{VAZQUEZ2017}, the ongoing MME is the most worrying and widespread threat to \textit{P. nobilis} throughout the Mediterranean Sea. As a consequence, the species has been declared as critically endangered \citep{IUCN}. Although different aetiological agents have been proposed, including Mycobacteria and other bacteria \citep{Carella2019, Saric2020, Scarpa2020}, there is evidence that the main cause of this mortality is the protozoan \textit{Haplosporidium pinnae} \citep{DARRIBA201714, CATANESE20189, Box2020}, a new species that belongs to the genus Haplosporidium, one of the four genera of the protist order Haplosporida, where it has been found that other Haplosporidian parasites are behind the extensive mortality of several oyster species \citep{Burreson2004, Arzul2015}. Life stages include uninucleate and binucleate cells, plasmodia, and spores. A group of experts following up the event predicted a high risk that the disease would be spread by marine currents through the Mediterranean basin, which could cause the extinction of the species \citep{Cabanellas2019} as it is endemic. This has helped to better understand the spread of the disease, and identified surface currents as the main factor influencing local dispersion, whereas environmental factors influence the disease expression, which seems to be favoured by temperatures above $\SI{13.5}\,{}^o$C and a salinity range between $36.5$ and $\SI{39.7}{psu}$.
    
   In summary, we introduce and study in detail the properties of the mean-field version of a general compartmental model to study marine epidemics for bivalve populations, namely passive filtering sessile invertebrate hosts infected through waterborne parasites.
   There are two main hypotheses, the first one that a population level description (i.e. without the consideration of spatial effects) is able to describe well the dynamics of the epidemic in a relatively dense population in small bounded regions.
    A second assumption is that the host becomes infected with some probability, but that there is not a critical parasite load in the infection process. 
   After presenting the full SIRP model, then three different reductions are discussed, one exact, an approximate reduction of the former and a third reduction based on a timescale approximation. The study is closed with a validation with the available experimental data for the infection process of  \textit{Pinna nobilis} kept in tanks. We wish to point out, that being a highly endangered and protected species, the reported data correspond to an \textit{unintended experiment} that cannot be repeated, and maybe these data represent the only opportunity to estimate the fundamental parameters of the model. In addition, the setup in which the \textit{Pinna nobilis} were kept in tanks, represent themselves the ideal implementation of the conditions under which the mean-field model SIRP is valid.

\section{The SIRP model} \label{sec:model}

\subsection{Model structure and initial considerations} \label{subsec:modtruct}

    In this work we analyse the SIRP model, a deterministic multi-compartmental mean-field model,
    continuous in time and unstructured in spatial or age terms,
    to study infection in bivalve populations. In particular, we stress that the model as it is written describes spatially-homogeneous populations.
    Compartmental models are the
    most frequently used class of models in terrestrial epidemiology
    \citep{DiekmannBook}, and originated in the classic study of SIR models
    by Kermack and McKendrick \citep{McKendrick}. The use of compartmental
    models in the study of infectious processes in marine systems is quite
    rare until very recently \citep{Harvell2004}. As already advanced in 
    \cref{sec:Introduction}, there are relevant features in the description of epidemic processes in
    marine ecosystems that are different with respect to the case of terrestrial ecosystems \citep{MCCALLUM_intro},
    and their study in marine environments is dominated so far by so called proliferation 
    models \citep{Powell2015}, which do not address the transmission of the 
    pathogen. See \citep{article_SIP} for a discussion of several compartment 
    models for the study of marine epizootics.

    Compartmental models of diseases in terrestrial ecosystems caused by microparasites (i.e. viruses, bacteria 
    and protozoans) do not consider a compartment to describe the 
    dynamics of the parasite \citep{May1979}, describing  just the different stages of the host. Infection typically occurs in $2$ ways: i) as a contact process, in which the microparasite is transmitted directly from a an infected host, I, by contact or through air in close proximity, to a susceptible host, S; ii) through a vector, that has acquired the microparasite by biting an infected host, I, and passes the microparasite to a susceptible host, S. In the first case one can describe the infection process through some probability that the individuals come close, while in the second it is very relevant to
    describe the vector mobility, and at least $2$ compartments, susceptible and infected vector, are typically needed.
    Once the microparasite enters the host, it proliferates inside it, 
    so the infection process can be described by using compartments for susceptible individuals, S, 
    infected individuals, I, and possible exposed individuals, E.
    In particular, transmission in terrestrial sessile organisms (e.g. plants) is generically vector-borne.
    In the case of marine ecosystems, infection typically occurs through water-borne parasites,
    in particular in filter-feeders sessile organisms, while vector-transmitted diseases
    are much less frequent. Parasites may be transported by diffusion, sea currents, or even active motion (i.e. if they have flagella). In any case, infection between sessile hosts is not through direct contact, but instead through the production and excretion of parasites by infected individuals and the assimilation by filtering of parasites by a healthy (susceptible) host. So, parasites are produced and excreted to the marine medium, in which they stay infective until they become deactivated (i.e die) or are absorbed by hosts. In a way, in parasite transmitted marine diseases parasites have a dual role: they are not only agents that induce infection
    but also act as vectors that transmit disease from an immobile infected host to a susceptible one.
    
    The SIRP model is a general mean-field compartmental model to describe epidemic transmission through water-borne parasites, that we think is specially adequate to describe epidemic transmission in sparsely located passive filter feeders, like many bivalves.  We exclude the case of colonies in which individuals are in close proximity, e.g. mussels, corals, etc, in which direct contact could be relevant and should be included in the model. In the SIRP model hosts are described through $3$ different compartments, as in the SIR model, that represent different evolution stages of the disease: a susceptible class of healthy individuals that may contract the disease, $S$, an infected class of individuals that may pass the disease through excretion of the parasite, $I$, and a class of removed (namely dead) individuals, $R$, that cannot be infected any more and that cannot transmit the disease plus an extra compartment, $P$, for the parasite population in the medium. It is important to note that invertebrates do not develop long-term immunity in the mammalian sense \citep{Powell2015}, and so, no compartment of individuals ``recovered with immunity'' is considered. However, bivalves have a first line of defense with hemocytes being able to fight  parasites and reduce their internal population. Nevertheless, available evidence indicates that the number of individuals that can achieve a full recovery is usually small and can be neglected, and so it is not necessary to consider a process in which individuals in the I compartment return to the S compartment at some rate, like in the SIS model. Instead, the population’s long-term response to disease, when it occurs, is through natural selection for genotypes characterised by increased resistance to or tolerance for the pathogen. As already advanced, the SIRP model includes a
    fourth compartment that represents the parasite population in the water medium, whose population needs to be described explicitly. An explicit compartment allows to model the situation in which the population of parasites may evolve dynamically in time, although in Sec.~\ref{sec:fastslow} we will consider the case in which the parasite populations accommodates almost instantaneously to the infected host population, and the description of the parasite can be simplified.

    Infection occurs when the host enters in contact with the parasite in the marine medium. It involves a process of filtration of water by the bivalve, and although a detailed representation has been discussed in the literature \citep{BIDEGAIN_2016_2}, in the current SIRP model it is represented in an effective way.
    Infection is modelled through a nonlinear term, typical in compartmental models, 
but that now depends on the parasite population and not on the population of the infected compartment, $I$. In terrestrial epidemiology there are two alternative ways to represent infection \citep{MartchevaBook}; i) mass action incidence, in which infection grows as the population gets larger, $\beta S I$; ii) standard incidence, in which the infection is bounded as the population grows, $\beta S I/N$, where $N$ is the total (host) population. One must look at these two choices as limit cases, with the possibility that in reality the system is best described by an intermediate form, closer to one of the limit cases, for example the modified SIR model in \citep{Brauer1990} in which the infection term has $S+I$ in the denominator instead of the total population $N$, because the $R$ compartment is removed.
Modelling infection with an explicit representation of the parasite population encounters the same basic \textit{dilemma} about whether the incidence grows as the (host) population increases or is bounded. We will model infection as $\bar{\beta} S P$, where the two possibilities are equivalent to the two different incidences just mentioned: i) $\bar{\beta}=\beta$; ii) $\bar{\beta}=\beta/N$, where $\beta$, the disease transmission rate, is a \textit{constant}, that can depend at most on external parameters, like temperature and salinity, but not on the variables defining the model (populations of host compartments or parasites). The model is valid considering both types of incidence, and in the case study we will see which incidence seems more adequate in this case.

\subsection{General SIRP model} \label{subsec:SIRPmodel}
    
In this section we will write a mean-field compartmental model to describe epidemics of immobile (sessile) hosts in a marine medium through infection by a water-borne parasite. Being a mean-field model implies that the model is compartmental and does not include an explicit space dependence, and so, describes a well mixed system. The mean-field model describes a spatially homogeneous system, but we hope that it will be the basis for spatially inhomogeneous situations, by adding suitable terms accounting for the mobility of the parasite.

\begin{figure}[H]
        \centering
        \includegraphics[width=0.8\textwidth]{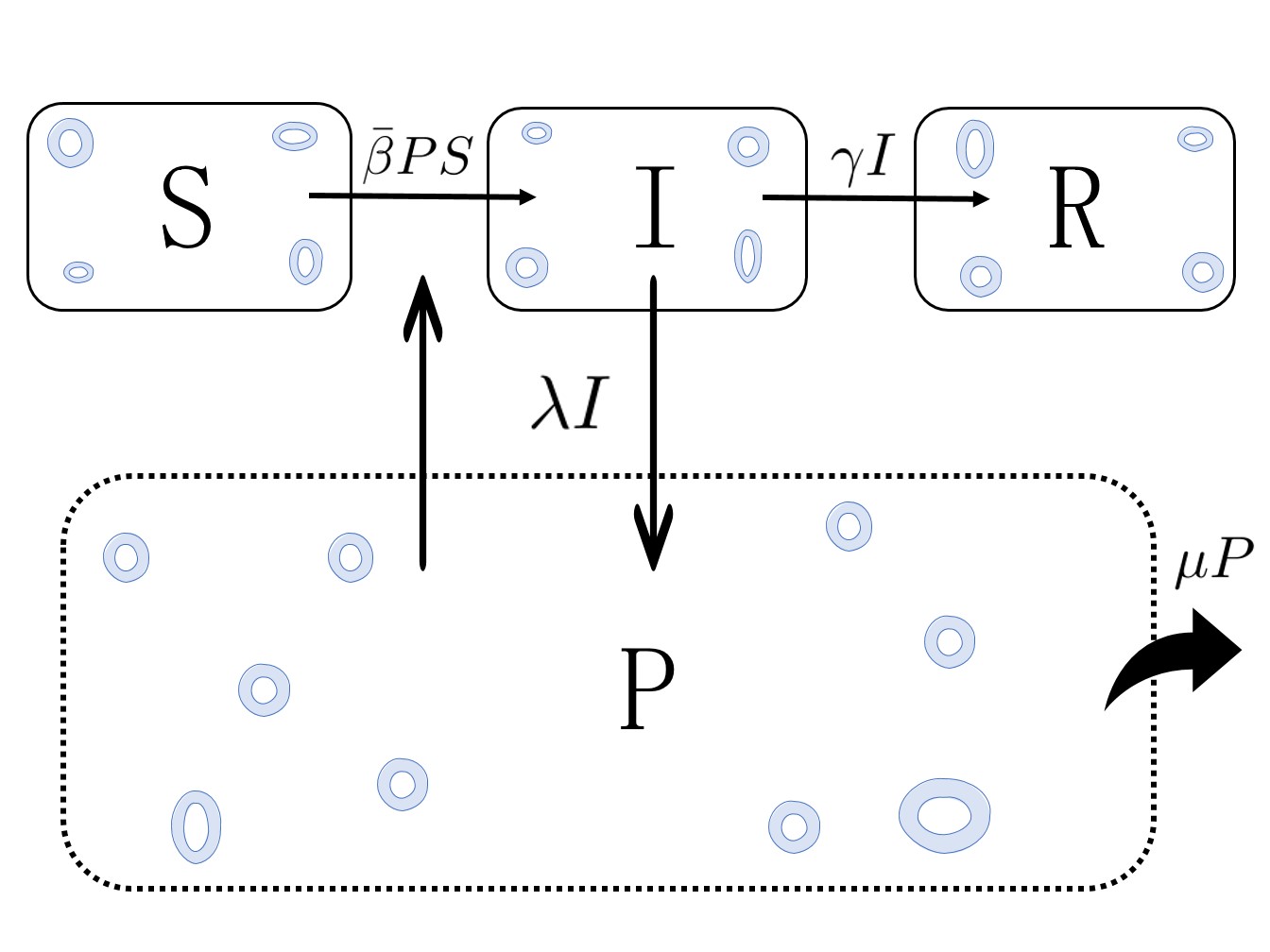}
        \caption{SIRP model flow diagram. The model variables are represented by capital letters: susceptible hosts ($S$), infected hosts ($I$), dead hosts ($R$) and the population of parasite ($P$). Arrows represent the processes in the model with their rates indicated next to them and blue rings represent parasites. The flow follows the scheme in \cref{eq:scheme_reinfection}, that leads to the system of differential equations in \cref{eq:SIRP}.}
        \label{fig: SIRP_scheme}
    \end{figure}

It is also assumed that the hosts become infected with some probability when exposed to a parasite, i.e., that there is not a critical parasite load needed for infection.
The model is defined according to the following reaction processes,
    \begin{equation}\label{eq:scheme_reinfection}
        S+P \stackrel{\bar{\beta}}{\rightarrow} I + \varnothing \quad I  \stackrel{\gamma}{\rightarrow} R \quad I \stackrel{\lambda}{\rightarrow} I+P \quad P \stackrel{\mu}{\rightarrow} \varnothing \ ,
    \end{equation}   
    which is graphically summarised in \cref{fig: SIRP_scheme}.
    
According to the scheme in \cref{fig: SIRP_scheme}, we consider the host in 3 possible states: susceptible, $S$, infected by the parasite, $I$ and removed (dead), $R$. Then we introduce the parasite population in the medium (water), $P$. In the model, the $\bar{\beta}$, the disease transmission rate parameter regulates the infection rate of susceptible hosts and accounts, among other mechanisms, for the parasite intake rate,
$\gamma$ the mortality of infected hosts, being the inverse of the typical mean time for an infected host to die; $\lambda$ the production rate of parasites from infected hosts, and $\mu$ the inverse of the typical life time of the parasite. $\mu$ can be related to several processes, like biological deactivation (or survival time) or other general losses, like dilution due to renewal of water in a closed experiment, natural losses in open ecosystems or absorption by other filter feeders.
We are not considering the possibility of spontaneous parasite gain, i.e. immigration, in this version of the model. A summary of the model parameters can be found in \cref{tab:parameters}.
We do not consider vital dynamics for the hosts, and this implies that the sum of the $3$ host subpopulations is constant, $N=S+I+R$, as the time scale of the disease evolution is much faster than the typical life cycle of fan mussels. 
The model is similar to the SIP presented in \citep{article_SIP}, except for an extra term in $\dot{P}$, $-\bar{\beta}PS$, accounting for the fact that when a parasite infects a host it is absorbed by it. The conditions under which the SIRP model can be simplified to the SIP model are discussed in \cref{sec:apprexred}.

    In order to build the deterministic model we consider that the population is large enough to neglect fluctuations and that it is well mixed, so that spatial effects can be neglected.
    In this situation, we consider the infection process to be proportional to the number of parasites in the medium, so that the average number of contacts between susceptible fan mussels and the average parasite population is given by $PS$, and, thus, the change in the number of susceptible fan mussels takes the form $\dot{S}=-\bar{\beta} PS$,  where the dot over a variable indicates a differentiation with respect to time: $\dot{S}=dS/dt$.
    
    \begin{table}[H]
        \centering
        \caption{Model parameters description}
        \begin{tabular}{cl:cl}
            \hline \hline
            Variable & Definition & Parameter & Definition \\ \hline
            $S$ & Susceptible host & $\beta$ & Disease transmission rate \\
            $I$ & Infected host & $\gamma$ & Host mortality rate \\
            $R$ & Removed host & $\lambda$ & Production rate of parasites by infected hosts  \\
            $P$ & Parasite in the medium & $\mu$ & Parasite deactivation/dilution rate \\ \hline \hline
        \end{tabular}
        \label{tab:parameters}
    \end{table}
      
    Following this argumentation, the scheme in \cref{eq:scheme_reinfection} and \cref{fig: SIRP_scheme}, one can write the evolution equations of the SIRP model,
    \begin{equation}\label{eq:SIRP}
        \begin{aligned}
            \dot{S} &=-\bar{\beta} P S \\
            \dot{I} &=\bar{\beta} P S-\gamma I \\
            \dot{R} &=\gamma I \\
            \dot{P} &=\lambda I-\bar{\beta} P S-\mu P \ .
        \end{aligned}
    \end{equation}
    Model (\cref{eq:SIRP})  \textit{lives} in the $4$-dimensional (S,I,R,P) phase space, representing the variables 
    the populations of individuals in the susceptible, infected and removed host compartments and of parasites, respectively. These variables could be redefined so that S, I and R represent proportions of hosts in each compartment and P the population of parasites per host. 
    
    The fixed points of \cref{eq:SIRP} are determined by the conditions\footnote{We do not consider the trivial fixed point $S=I=R=P=0$ that would imply $N=0$ and $P=0$ at all time.} 
    $I=P=0$, to be fulfilled simultaneously. 
    We will study the stability of the fixed point defined by $S(0)$, $I(0)=P(0)=0$ and $R(0)=N-S(0)$.
    A linear stability analysis of this fixed point reveals that it has two null eigenvalues, that stem from the condition $N=S+I+R$ and the conserved quantity of \ref{app:P_exact}. 
    The first condition, $S+I+R=N$, implies that it is enough to consider two of the host populations, e.g. $S$ and $I$, as the third one can be obtained from the other two. The implications of the conserved quantity reported in \ref{app:P_exact} are more subtle, as it implies that fixed points are not isolated, as it happens in ordinary dissipative dynamical systems, and there is an infinite number (a line of) fixed points for the final state of the epidemic, depending on the initial conditions. This also implies that the phase space is foliated by the conserved quantity, $C$ of \cref{eq:conservedquantity}, and every initial condition, $S_0$, with a different value of $C$ leads to a different asymptotic condition, $S_{\infty}$, just as shown in \citep{Murray_book} for the SIR model (cf. Fig. 10.1 in \textit{op. cit.}).
    The third eigenvalue, that is the largest of the two non-zero eigenvalues, can be positive if $\beta S_0 \lambda>\gamma(\beta S_0+\mu)$ and negative if the inequality is reversed, defining the conditional stability of the fixed point. The fourth eigenvalue is always negative and all the eigenvalues are always real (cf. \ref{app:linstabfp}).
    The instability of the fixed point along the third eigenvalue drives the beginning of the epidemic.

    An extremely important result in epidemiology is the so-called \textit{basic reproduction number}, $R_0$, a dimensionless number which represents the number of secondary infections produced by a primary infection in a fully susceptible population. $R_0=1$ defines the threshold for epidemic propagation: an epidemic will occur when $R_0>1$, and the number of infected individuals will grow, at an exponential rate in the early phases of the epidemic \citep{Castro2020}, while if $R_0<1$ the infection will wane naturally.
    This quantity can be formally obtained making use of the Next Generation Matrix (NGM) method \citep{Theory_next_gen_matrix, Diekmann2010}. Applying this formal method to our system of ordinary differential equations (ODE's) one obtains the following relation for the basic reproduction number (cf. \ref{app:NGM}),
    \begin{equation}\label{eq:R_0_SIRP}
        R_0=\frac{\lambda}{\gamma\parentesi{1+\displaystyle\frac{\mu}{\bar{\beta} S(0)}}} \ .
    \end{equation}
     
    The threshold condition provided by $R_0$ (\cref{eq:R_0_SIRP}) is equivalent
    to the linear stability condition for the third eigenvalue of the initial, pre-epidemic, fixed point, as $\bar{\beta} S(0) \lambda>\gamma(\beta S(0)+\mu)$
    implies that this eigenvalue is positive and the disease-free equilibrium state unstable being this
    equivalent to $R_0>1$ (cf. \ref{app:linstabfp}).
    Thus, if $R_0>1$ the fixed point is unstable, and an epidemic will ensure if infected hosts, $I$, or parasites, $P$ appear in the system. An epidemic will propagate until the system reaches an stable fixed point, that signals the end of the epidemic (cf. \ref{app:linstabfp}).

\subsection{Model reduction} \label{sec:reduction}

    The SIRP model lives in a $4$-dimensional phase space and depends on $4$ parameters, what makes difficult to confront it with experimental data. Thus, we will discuss here 
    three alternative ways of reducing the model. The first involves an exact reduction of the model, based on the conserved quantity derived in \ref{app:P_exact}. The second reduction consists of an approximation to the previous exact reduction, that turns out to be equivalent to an exact reduction of a slightly simplified model (without the $-\bar{\beta}SP$ term in the equation of $\dot{P}$). The third one is based on an approximation valid if the system parameters fulfil certain conditions. 

\subsubsection{Exact reduction of the SIRP model} \label{sec:exactred}
    
    From the conserved quantity derived in  \ref{app:P_exact}, it is possible to write the parasite population in the SIRP model as a function of the host states as follows,
    \begin{equation}\label{eq:P_exact}
         P(S,I)=-\frac{\lambda}{\gamma}\parentesi{S+I} + \frac{\mu}{\bar{\beta}}\ln(S) + S + C(0) \ ,
    \end{equation}
    where $C(0)=P(0)+\displaystyle\frac{\lambda}{\gamma}\parentesi{S(0)+I(0)} - \frac{\mu}{\bar{\beta}}\ln(S(0)) - S(0)$.\\
    
    Substituting \cref{eq:P_exact} into the general SIRP model of \cref{eq:SIRP} we obtain the following nonstandard SIR model,
    \begin{equation}\label{eq:SIR_exact}
        \begin{aligned}
            \dot{S} &= \frac{\lambda\bar{\beta}}{\gamma}S\parentesi{S+I} - \mu S\ln(S) - \bar{\beta} S^2 - S\bar{\beta} C(0) \\ 
            \dot{I} &= -\frac{\lambda\bar{\beta}}{\gamma}S\parentesi{S+I} + \mu S\ln(S) + \bar{\beta} S^2 + S\bar{\beta} C(0) -\gamma I \\
            \dot{R} &=\gamma I \ .
        \end{aligned}
    \end{equation}
    
    Although using the conserved quantity yields an exact reduction from a 4D dynamical system to a 3D one, the number of independent parameters and initial conditions remain unchanged, i.e. they still depend on $4$ parameters and $4$ initial conditions. Thus, although useful, (\ref{eq:SIR_exact}) is not ideal when trying to fit experimental data, and this is the reason for trying a further approximation to \cref{eq:SIR_exact} to be discussed next.
   
    \subsubsection{Further approximation to the exact reduction} \label{sec:exactredapp}
    
    A further approximation to \cref{sec:exactred}, that is less restrictive and expected to be valid in a broader parameter range than \cref{sec:fastslow} is possible. This approximation reduces the number of free parameters by one, what is useful in fitting available data. 
    The approximation consists of neglecting the
    $S$ term in \cref{eq:P_exact}, what is possible if $\lambda/\gamma\gg 1$ and also $\mu\ln N/(\bar{\beta} N)\gg 1$, as $S(t)$ decreases monotonically with time and is, at most, $N$ at the initial time. Interestingly, this approximation is equivalent to the simplification of the equation for $\dot{P}$ in (\ref{eq:SIRP}) so that the $-\bar{\beta}SP$ is skipped, what yields exactly the SIP model of Ref. \citep{article_SIP}. This reduced model has an exact conserved quantity, $\mathcal{C}$, that differs from that of the SIRP model in the linear $S$ term (cf. \ref{app:P_exact}).
    Using this approximation one can write, 
    \begin{equation}\label{eq:SIR_exact_reduced}
        \begin{aligned}
            \dot{S} &=\frac{\lambda'}{\gamma}S(S+I)-\mu S\ln(S)-S\bar{\mathcal{C}}(0) \\
            \dot{I} &=-\frac{\lambda'}{\gamma}S(S+I)+\mu S\ln(S)+S\bar{\mathcal{C}}(0)-\gamma I \\
            \dot{R} &=\gamma I \ ,
        \end{aligned}
    \end{equation}
    where $\lambda'=\lambda\bar{\beta}$ and $\bar{\mathcal{C}}(0)=\bar{\beta} (P(0)+\lambda/\gamma(S(0)+I(0)-\mu/\bar{\beta}\ln S(0)))=\bar{\beta} \mathcal{C}(0)$ is a redefinition of the conserved quantity of the SIP model \cref{eq:conservedquantitySIP}, $\mathcal{C}(0)$, a constant, such that it absorbs $\bar{\beta}$ and all initial conditions of the model. The result is that \cref{eq:SIR_exact_reduced} depends on $3$ parameters and $1$ constant, compared to \cref{eq:SIR_exact} that depends on $4$ parameters,
    facilitating, thus, the use of the model to fit experimental data.
    
     \subsubsection{Model reduction through fast-slow separation} \label{sec:fastslow}
    
    The third 
    reduction of the $4$-D dynamical model \cref{eq:SIRP} makes the assumption that
    the time scale in which the parasite population changes are faster than the one corresponding to the host. This means that pathogen deactivation in the medium must be faster than host mortality. In terms of the rates associated to each of these processes, this means $\mu>\gamma$. Taking $\mu$ as common factor in $\dot{P}$ one can write,
    \begin{equation}
        \epsilon \dot{P} =\lambda I/\mu-\bar{\beta}SP/\mu-P 
    \end{equation}
    where $\epsilon=1/\mu$ is small, as $\mu$ is large. If furthermore $\mu\gg\bar{\beta}N$ and  $\lambda\gg \beta P$ one arrives to,
    \begin{equation}\label{eq:P_approx}
        P\approx \frac{\lambda}{\mu} I \ .
    \end{equation}
    
    Under this approximation the slow subsystem can be written,
    \begin{equation}\label{eq:SIRslow}
        \begin{aligned}
            \dot{S} &=-\beta' I S \\
            \dot{I} &=(\beta' S-\gamma) I \\
            \dot{R} &=\gamma I\ ,
        \end{aligned}
    \end{equation}
    that is equivalent to the classical SIR model with $\beta'=\bar{\beta} \lambda/\mu$ instead of the infection rate $\bar{\beta}$. The reduced $3$-D model \cref{eq:SIRslow} from the original $4$-D SIRP model \cref{eq:SIRP} depends on $2$ parameters instead of $4$ as the original model had and $1$ initial condition, e.g. $I(0)=N-S(0)$ if $R(0)=0$, and is much more amenable to be applied to the analysis of experimental data, as shown in \cref{sec:validation}. 
    Furthermore, $\gamma$ could be eliminated through a time rescaling, $t\rightarrow=t'=\gamma t$ with a redefinition of $\beta'\rightarrow\beta''=\beta'/\gamma=\beta\lambda/(\mu\gamma)$, leaving the model as a function of a single effective parameter. However, we will keep both $\beta'$ and $\gamma$ for convenience when fitting the model to experimental data in \cref{sec:validation}, as we would need to know anyhow $\gamma$ in order to analyse the experimental data. The validity of this approximation is checked numerically in \cref{sec:numericalanalysis}.

\section{Numerical analysis of the model} \label{sec:numericalanalysis}

    Due to the impossibility of solving the SIRP model analytically,
    in the present section we perform a numerical characterisation of the 
    model\footnote{All numerical simulations of the dynamical system \cref{eq:SIRP} have been carried out using a Runge-Kutta $4$th order method, with a temporal step $\Delta t=0.001$. Numerically stable results are obtain with $\Delta t\le 0.01$.}.
    Moreover, we show the validity range of the performed approximations to reduce the SIRP model to an effective SIR model. We start our numerical analysis by investigating the relative influence of the model parameters on some epidemiological quantities of interest: the basic reproduction number ($R_0$), related to the existence of an epidemic outbreak, continuing with the final state of the epidemic, given by the final number of dead individuals ($R(\infty)$) and the maximum of infected individuals ($I_{\textrm{max}}$) together with the time at which it occurs ($t_{\textrm{max}}$). 

    In order to identify the most influential parameters of our model Sensitivity Analysis (SA) will be performed. SA can be divided into two classes: Local Sensitivity Analysis (LSA) and Global Sensitivity Analysis (GSA). LSA represents the assessment of the local impact of input factors variation on model response by concentrating on the sensitivity in the vicinity of a set of factor values. Such sensitivity is often evaluated through gradients or partial derivatives of the output functions at these factor values, such that other input factors are kept constant.
    Since epidemic models exhibit a threshold behaviour, controlled by the dimensionless quantity
    $R_0$, it is relevant to study its robustness with respect to small perturbations by means of the LSA explained above, as its analytical expression is known.
 
    On the other hand, GSA will be applied to study the influence of the parameters in the final state of the epidemic and the epidemic peak by exploring a large domain of the parameter space. In turn, GSA is the process of apportioning the uncertainty in outputs to the uncertainty in each input factor over their entire range of interest. A sensitivity analysis is considered to be global when all the input factors are varied simultaneously and the sensitivity is evaluated over the entire range of each input factor, in clear contrast to LSA. Within GSA, first order indices are a measure of the contribution to the output variance given by the variation of the parameter alone averaged over variations in other input parameters while second order indices take into account first order interactions between parameters. While LSA is carried out analytically (if exact expressions are available), GSA is a purely numerical approach. Further mathematical details on Sensitivity Analysis can be found in \ref{app:sensanal}.
    
    For all the sensitivity analysis performed in the following sections, and in order to avoid ambiguities associated to the definition of $\bar{\beta}$ as a function of $N$, we assume $N=1$, so that both possible incidences yield $\bar{\beta}=\beta$ and the numerical results are equivalent.
    
\subsection{The basic reproduction number $R_0$}

    To study the relevance of parameters involved in an epidemic outbreak a LSA was performed. We analyse the local sensitivity of $R_0$ through the normalised sensitivity index, so that the function $F(\vec{p})$ of \cref{eq:sensitivity_analysis} is substituted by the analytical expression of $R_0$, \cref{eq:R_0_SIRP}. 

\begin{figure}[H]
        \centering
        \includegraphics[width=1\textwidth]{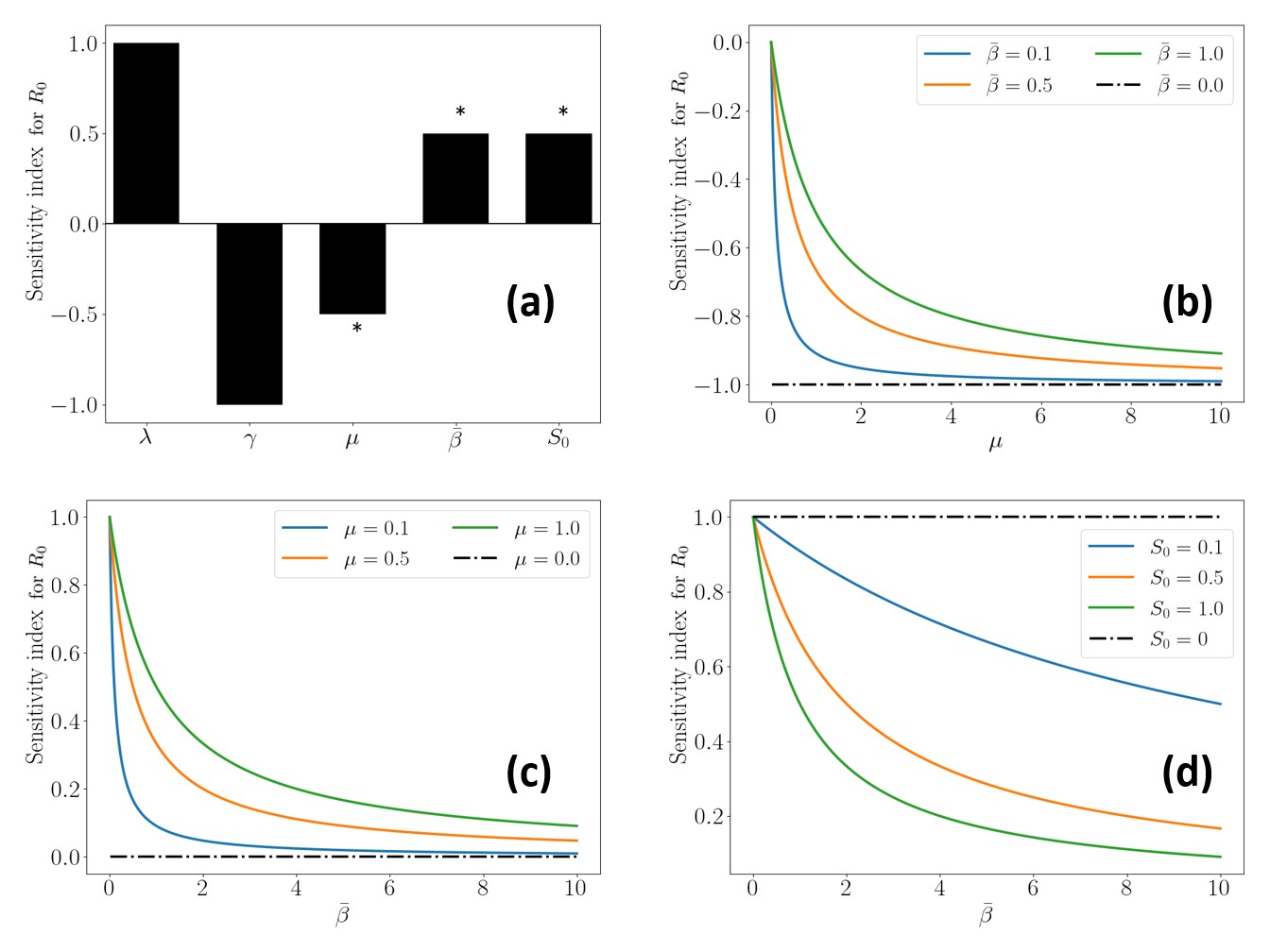}
        \caption{Panel (a): Local sensitivity analysis of $R_0$ for the  baseline parameters $\lambda=1$, $\gamma=1$ and $\mu=\bar{\beta}=S_0=1$. The asterisks mark parameters for which the sensitivity index is not constant, depending on, at least, another parameter. Panels (b-d): Local sensitivity analysis of $R_0$ with respect to parameters with an asterisk, showing the different dependence with a second parameter and the effect on the varying sensitivity index.}
        \label{fig:Local_Sensitivity_Analysis_R0}
    \end{figure}

    \cref{fig:Local_Sensitivity_Analysis_R0}(a) shows the sensitivity index for $R_0$ for specific baseline parameters, where $\lambda$, $\bar{\beta}$ and $S_0$ contribute to increase the basic reproduction number while $\alpha$, $\gamma$ and $\mu$ contribute to decrease it, as expected. Moreover, we can see that $\lambda$ and $\gamma$ are the most influential parameters while $\mu$, $\bar{\beta}$ and $S_0$ depend on each other. These dependencies cause varying influences on $R_0$, which are fully depicted in panels \cref{fig:Local_Sensitivity_Analysis_R0}(b-d). It can be seen that the influence of $\bar{\beta}$ increases with the increase of $\mu$ and the decrease of $S_0$. Similarly, the importance of $S_0$ increases with $\mu$ and decreases with $\bar{\beta}$. On the other hand, the impact of $\mu$ increases with the decrease of both $S_0$ or $\bar{\beta}$.

        \begin{figure}[H]
        \centering
        \includegraphics[width=0.7\textwidth]{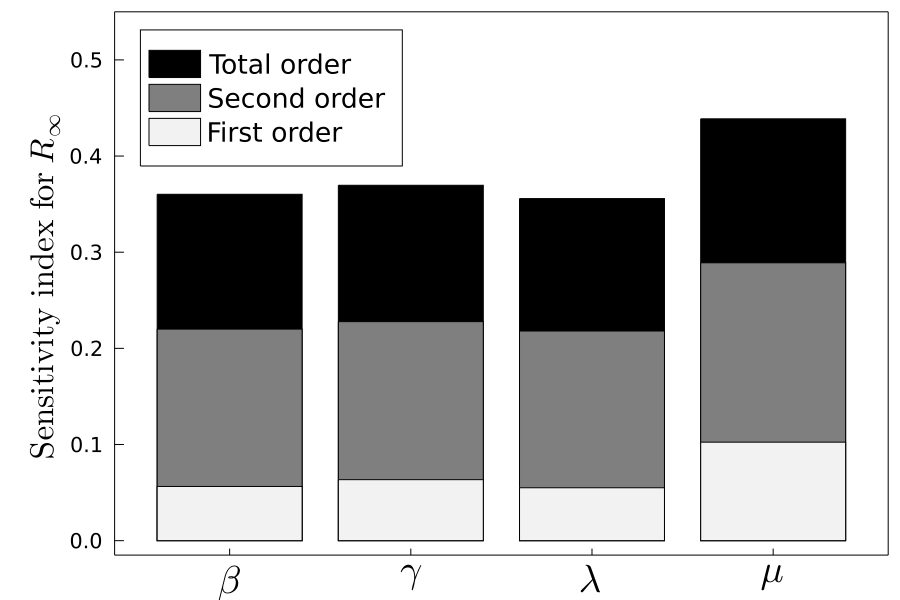}
        \caption{Sensitivity indices (LSA) for the final number of dead individuals ($R_\infty$) for each one of the indicated parameters. The black bars represent the total order indices of sensitivity while white (grey) colour represents the contribution of the first (second) order indices.}
        \label{fig: GSA_R_inf}
    \end{figure}

    \subsection{Final state of the epidemic}

    Another important quantity in epidemiology is the final state of the epidemic, which can be characterised by the final number of dead individuals, $R_\infty$. Within our general SIRP model it is not possible to find an analytical expression of $R(t)$ so that we need to tackle the problem numerically. To this end, we perform GSA for the final number of dead individuals in order to determine the most influential parameters for this quantity. In particular, we apply the Sobol method, discussed in \ref{app:sensanal}.
    The Confidence Interval, CI, obtained in our 
    study is less than 1\% of the index value, indicating a very high accuracy, therefore it is not shown in the figures. The results of the explained procedure are shown in \cref{fig: GSA_R_inf}, where the total order (black), first order (white) and second order (gray) sensitivity indices for each of the model parameters are detailed. It can be observed that $\mu$ has a slightly greater influence than the other parameters with respect to the final number of dead individuals. Note that the second order indices are larger than the first order ones for all the parameters, which indicates a high influence of the nonlinearities in our model, at least for the particular quantity under study.
 
  \subsection{Maximum of infected individuals}

\begin{figure}[H]
        \centering
        \subfigure[Epidemic peak]{\includegraphics[width=0.49\textwidth]{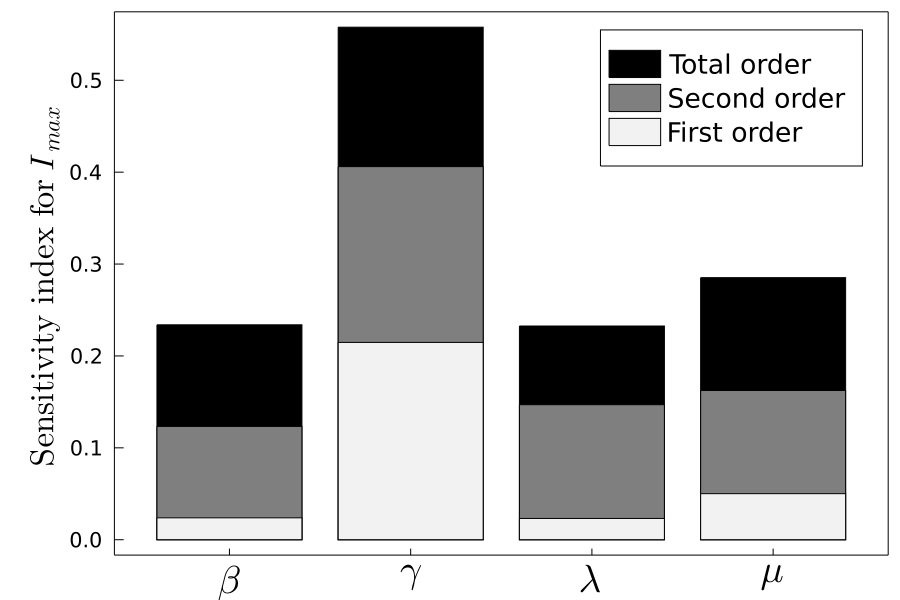}}
        \subfigure[Time of epidemic peak]{\includegraphics[width=0.49\textwidth]{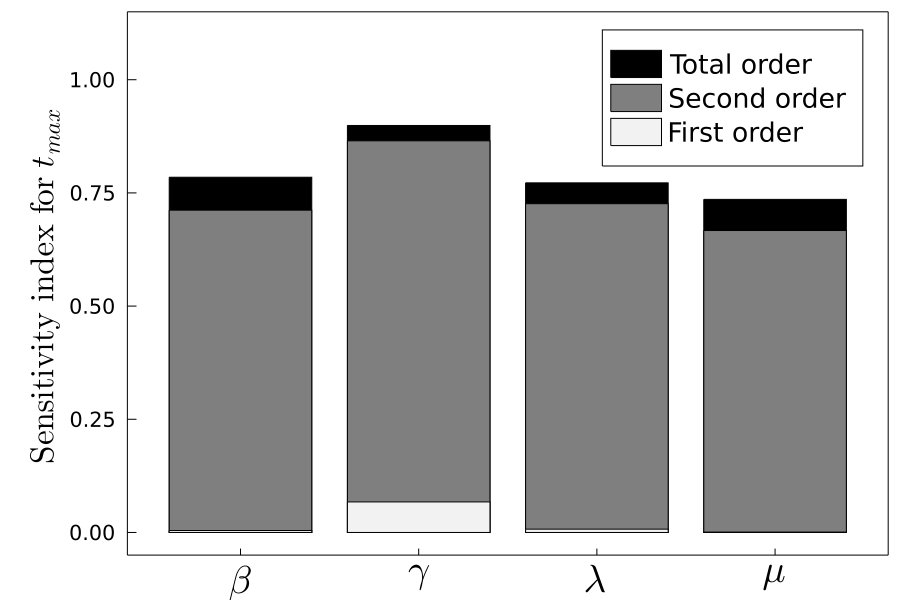}}
        \caption{Global sensitivity analysis for the maximum of infected individuals $I_{\textrm{max}}$ (a) and its time occurrence $t_{\textrm{max}}$ (b). The black bars represent sensitivity at all orders,
        while white (grey) colour represents the contribution of the first (second) order indices.}
        \label{fig: GS_analysis}
    \end{figure}

    A GSA of the maximum number of infected individuals, $I_{\textrm{max}}$ and the time it occurs, $t_{\textrm{max}}$ is performed to study the influence of the model parameters regarding these quantities. In this case, \cref{fig: GS_analysis}, $\gamma$ has greater influence in the epidemic peak than any of the other parameters, while for the time at which the peak takes place, all the parameters have basically the same influence. Again, the second order indices (the first order interactions between parameters) account for most of parameter sensitivity, in particular in the time of the epidemic peak, indicating the high degree of nonlinearity of this effect.

\subsection{Numerical verification of the fast-slow approximation} 

\begin{figure}[H]
        \centering
        \includegraphics[width=1\textwidth]{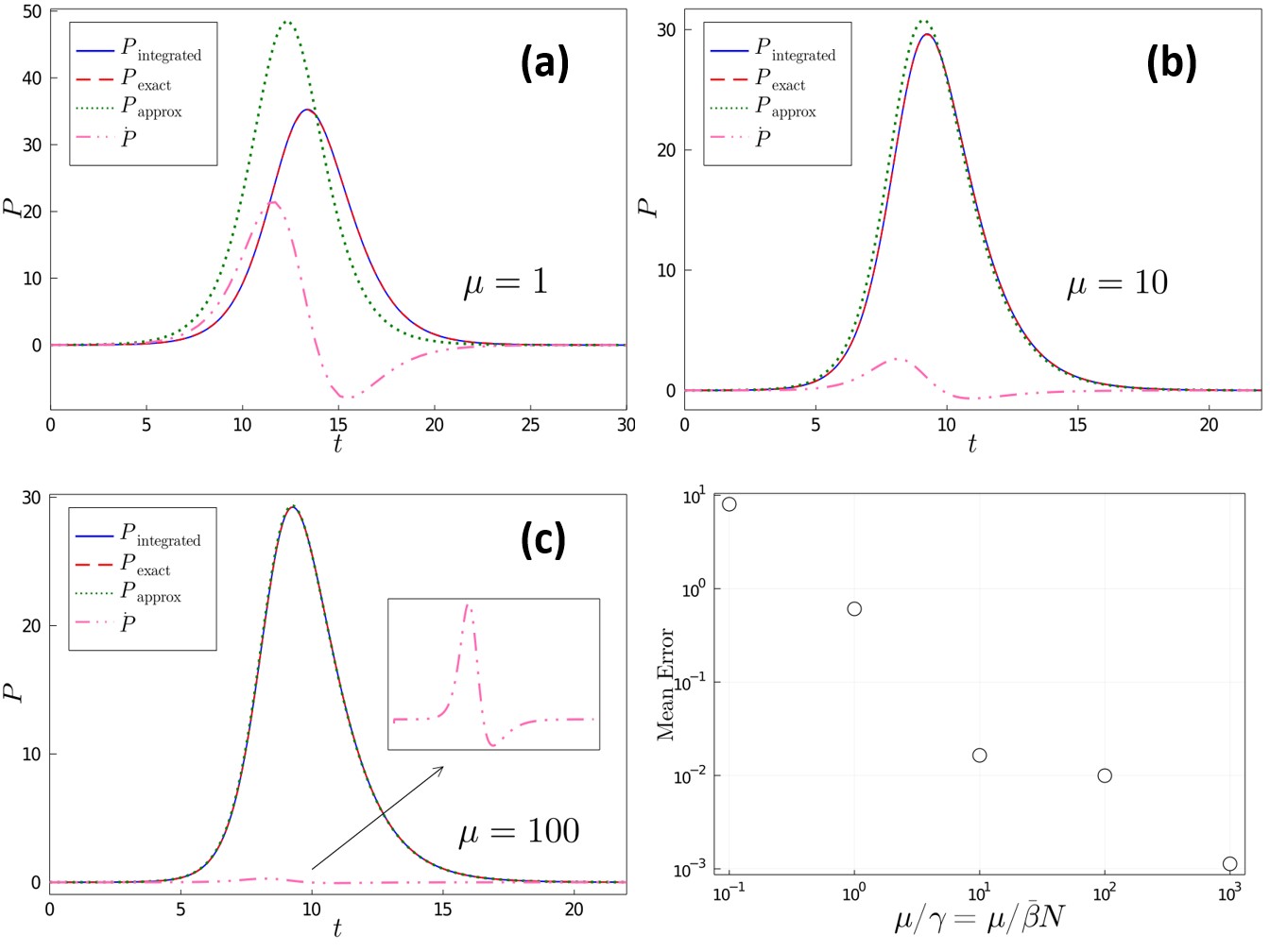}
        \caption{Numerical check of the approximate expression for the pathogen concentration, (\cref{eq:P_approx}), $\bar{\beta}=1/50$ and $\gamma=1$: (a) $\mu=1$; (b) $\mu=10$; (c) $\mu=100$, while $\lambda$ is varied to keep $R_0=2.5$, defined in (\cref{eq:R_0_SIRP}) (with $S_0=N=50$), i.e., $\lambda=5, 27.5, 252.5$ respectively for (a)-(b)-(c), respectively.
        The blue solid line represents the numerically integrated quantity, the red dashed line (superimposed to the blue solid one as they are identical) is the exact solution for this quantity, (\cref{eq:P_exact}) and the green dotted line accounts for the approximate expression from the timescale separation (\cref{eq:P_approx}). The dash-dotted pink line represents the derivative of $P$, $\dot{P}$, in the scaled time frame. Panel (d): Mean error between the approximate and exact solutions for increasing $\mu/\gamma=\mu/\bar{\beta}$.}
        \label{fig:P_comparison}
    \end{figure}

    The parasite concentration approximation, based on a timescale separation discussed in \cref{sec:fastslow}, is now verified by computational means. The verification was performed using both mass action and standard incidence, but for the sake of simplicity we show only the results for the standard incidence case. Worth is to say that, mathematically, changing from standard incidence to mass action involves only a rescaling of the $\beta$ parameter, so that the numerical results are exactly the same. \cref{fig:P_comparison} contains a comparison for $3$ different values of the parasite deactivation rate, $\mu$. It can be seen that the approximation is poor when $\mu\sim \gamma,\bar{\beta}N$ \cref{fig:P_comparison}(a), as it could be expected. On the other hand, the approximation is quite good when $\mu$ is one order of magnitude larger than $\gamma$ and $\bar{\beta}N$ \cref{fig:P_comparison}(b), while it is extremely accurate when $\mu$ is two orders of magnitude larger than $\gamma,\bar{\beta}N$, \cref{fig:P_comparison}(c). The figure also shows the numerical value of $\dot{P}$ (pink dashdot), and it can be checked how it becomes smaller as $\mu$ increases compared to $\gamma,\bar{\beta}N$, justifying the timescale separation of \cref{sec:fastslow}. Finally, \cref{fig:P_comparison}(a-c) also shows (dashed red line) the analytical value for $P(S,I)$ derived in \cref{eq:PSI_exact}, that matches perfectly the result of the numerical integration of \cref{eq:SIRP}, as should be the case.
    
    \subsection{Numerical verification of the model approximation from the exact reduction} \label{sec:apprexred}
    
\begin{figure}[H]
        \centering
        \includegraphics[width=1\textwidth]{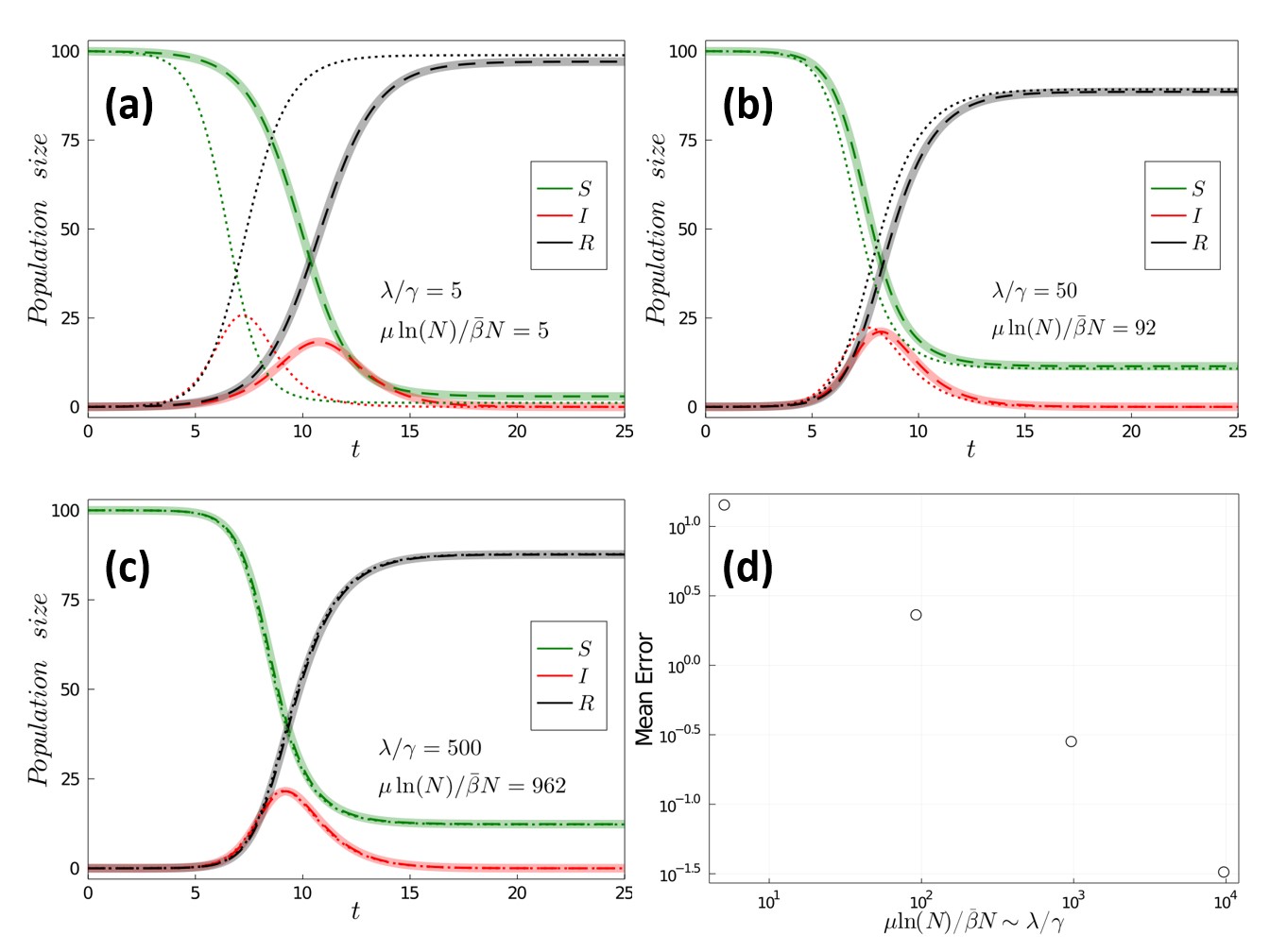}
        \caption{Numerical check of the exact model reduction along with the subsequent approximation shown in \cref{sec:exactred} with $N=100$, $\bar{\beta}=1/100$, $\gamma=1$, (a) $\lambda=5$, $\mu=1.1$; (b) $\lambda=50$, $\mu=20$; (c) $\lambda=500$, $\mu=209$. $R_0=2.38$ for all the panels. The solid semitransparent lines represent the original 4D model, the dashed lines the exact reduction and the dotted lines the approximate model from the exact reduction. Panel (d): Mean error between the approximate and exact solutions for increasing $\mu\ln(N)/\bar{\beta}N$ and $\lambda/\gamma$ while $R_0=2.38$ is kept constant.}
        \label{fig:verification_exact_reduction}
    \end{figure}

    The numerical verification was performed for both mass action and standard incidence, but for the sake of simplicity in \cref{fig:verification_exact_reduction} we show only the results for the standard incidence case. First, and as it should be because it is an exact result, the exact reduction of the SIRP model discussed in \cref{sec:exactred} matches perfectly the numerical results obtained from the full model for all possible parameter values,
    \cref{fig:verification_exact_reduction}(a-c).
    Regarding the approximation to the exact reduction, one can see 
    how the approximation converges to the exact solution as the parameters fulfil the conditions indicated in \cref{sec:exactredapp}, namely that both $\gamma/\lambda\gg 1$ and $\mu\log(N)/\bar{\beta N}\gg 1$, becoming very accurate if these ratios are larger than $1$ by two orders of magnitude or more (cf.  \cref{fig:verification_exact_reduction}(c)). We recall that in this case the SIRP model converges to the SIP model of \citep{article_SIP}. Conversely, the approximation is poor when any of these two ratios is of order $1$ ((cf.  \cref{fig:verification_exact_reduction}(a))), while \cref{fig:verification_exact_reduction}(b) presents the result in an intermediate case, in which the approximation is fair.

\section{Model validation with data of the \textit{Pinna nobilis} Mass Mortality Event}
\label{sec:validation}

    In this section, the general SIRP model is validated against collected data from the \textit{Pinna nobilis} Mass Mortality Event. As explained in \cref{sec:Introduction}, the disease is caused by the parasite \textit{Haplosporidium pinnae} and the hosts, \textit{P. nobilis}, are sessile bivalves endemic of the Mediterranean Sea. Thus, this epidemic is a perfect candidate to be described by the SIRP model. In the model, parasite production occurs only inside infected hosts, and parasites are released to the medium, either through their respiratory or digestive system. The simultaneous occurrence of the different possible stages of the parasite (uni- and bi-nucleate cells, multinucleate plasmodia, sporocysts and uninucleate spores) in the same host individual is not common among haplosporidans and makes \textit{H. pinnae}  different from previously known haplosporidan species \citep{CATANESE20189}. The occurrence of uni- and binucleate stages suggest possible direct transmission from infected to healthy fan mussels, as observed in \textit{B. ostreae} and \textit{B. exitiosa} \citep{hine1996ecology, Culloty2007, Audemard2014}. Additionally, the presence of spores (a dormant, resistant stage) could allow long persistence in the environment and the hypothetical involvement of an intermediate host as suggested for \textit{H. nelsoni} and \textit{H. costale} \citep{Andrews1984, haskin1988uncertainties, powell1999modeling}. While uninucleate cells are always detected in infected fan mussels, sporulation has been only detected sporadically \citep{CATANESE20189}. Thus, we assume that infection occurs mostly through uninucleate (or binucleate) cells by direct transmission (as the experimental observations in captivity point out, see \citep{March}). 
    We do not consider disease transmission through other stages. We do not consider spores, given the infrequent observation of spores and the current lack of experimental information about spore transmission (that could involve another intermediate host species). Regarding plasmodia and sporocyst stages, these stages are too large to be released through the epithelium. The distinction between uninucleate and binucleate cells seems unnecessary at this level of representation, as these phases only participate in parasite proliferation inside infected hosts, a process that we consider in an effective way. Finally, the evidence of the time course of the disease compared to the long life cycle of {\it P. nobilis\/} suggests host vital dynamics (i.e. recruitment (reproduction) and natural death) can be neglected.
    
    After an epidemic outbreak that took place in Portlligat, in the north east of Catalonia, $215$ \textit{Pinna nobilis} individuals were extracted from their natural medium in order to be preserved as a genetic reserve in several controlled water tanks of different institutions in Spain \citep{March}. The institutions that participated in this preservation effort were IFAPA, IEO, IRTA, IMEDMAR-UCV and Oceanogr\`afic of Valencia. The original idea was to rescue the individuals before infection, however, the subsequent evolution of the rescued \textit{Pinna nobilis} populations indicates that some individuals were already infected at the time of extraction (and/or in contact with some amount of the parasite transferred from sea water).
    This allowed the opportunity to use the data of the time evolution of the epidemic in the controlled water tanks, reported in \citep{March}, to evaluate the described SIRP model\footnote{Data use in this work with the purpose of validating and fitting parameters for the SIRP model have been taken from the Supplementary Information of \citep{March}}. The empirical data consists of the proportion of survivors as a function of time in the controlled water tanks with a temporal resolution of one month. Despite the fact that the temperature of the water in the tanks was controlled, it was sharply lowered in most of the tanks when mortality started to appear within the population, as a last effort to keep the rest of the population safe and alive, since keeping the temperature below  approximately $13.5\, {}^o$C is a known strategy to preserve \textit{Pinna nobilis} individuals as disease expression is minimal \citep{Cabanellas2019}. Fortunately, two of the tanks kept its temperature approximately constant during the full recorded time. This is the case of the tanks in IFAPA in Huelva and the Oceanogr\`afic of Valencia (OCE), both Spanish institutes. These water tanks have been selected to validate our model, maintaining constant temperatures of $14\, {}^o$C and $17\, {}^o$C, respectively. 

    \begin{figure}[H]
        \centering
        \includegraphics[width=1\textwidth]{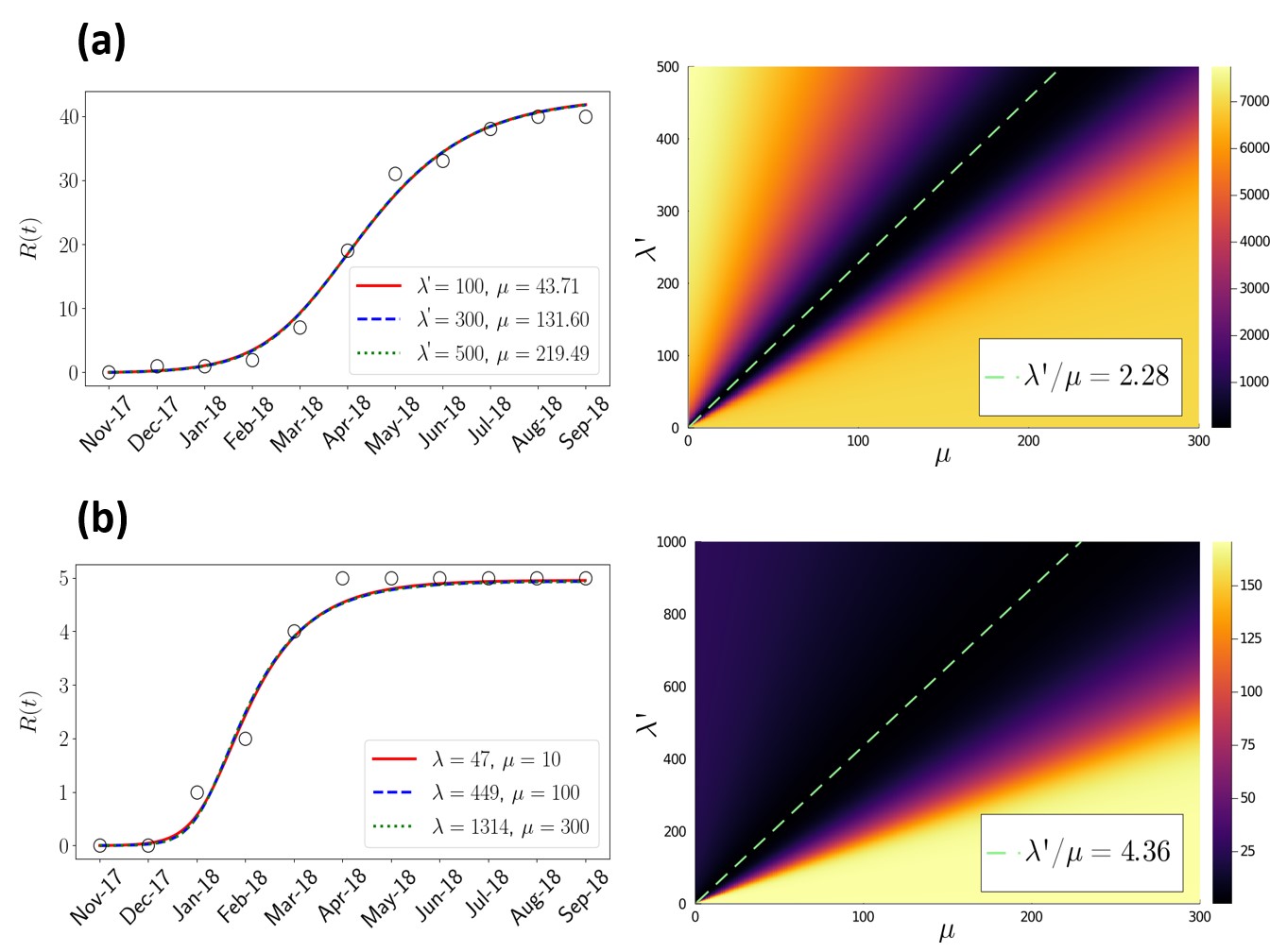}
        \caption{Parameter estimation for the approximation from the exact reduction of the SIRP model (\cref{eq:SIR_exact_reduced}) using data from IFAPA (panel (a)) and OCE (panel (b)) water tanks, at $\SI{14}{\degree C}$ and $\SI{17}{\degree C}$ respectively. Left figures represent several fits of the model to empirical data of the number of dead hosts ($R(t)$) using different optimal combinations of the parameters. Right figures are the RSS errors as a function of the input parameters, where the green dashed line represents the set of optimal combinations of the parameters with $RSS=60, 0.8$ for IFAPA and OCE, respectively.} 
        \label{fig: exact_SIR_fit}
    \end{figure}
   
    First we will fit the exact reduction of the SIRP model, assuming 
      $\mu\log(N)\gg\bar{\beta}N$ and $\lambda/\gamma\gg 1$ as discussed in \cref{sec:exactredapp}, namely \cref{eq:SIR_exact_reduced}.
    This reduced model depends on three parameters ($\lambda'$, $\mu$, $\gamma$) and one constant, $\bar{\mathcal{C}}(0)$, cf. \cref{sec:exactredapp}, that is related to the initial conditions of the model. The order of magnitude of the mortality rate can be deduced from data, with an estimate value of $\gamma\approx\SI{1}{month^{-1}}$. We fix this parameter in order to give some biological information to our model prior to the computational fit. We focus on the $R$ compartment, as it can be retrieved directly from data in \citep{March}\footnote{The number of dead individuals can be obtained as $R=N-S$, where $S$ is the population of survivors and $N$ is the total number of individuals in the tanks, 
    $50$ (IFAPA) and $5$ (Oceanogr\`afic), respectively}.
    We use a box-constrained variant\footnote{We constrain the optimisation because the unconstrained optimisation to the full range of the parameters, i.e, from $0$ to $\infty$ is not practical.} of the well known BFGS optimisation algorithm \citep{BFGS} with a common $\textrm{L}2$ loss function, also known as Residual Sum of Squares (RSS)\footnote{The algorithm is implemented within the Julia high-level programming language \citep{julia} using the DifferentialEquations.jl package \citep{DifferentialEquations.jl}.}. By running this algorithm one observes that the optimal parameters tend to be the ones in the boundary of the box-constrained parameter space. 
    Furthermore, if the box size is increased (or decreased) the optimal parameters continue to be in the boundary of the box-constrained parameter space.
    This indicates that there exist several parameter combinations that optimally fit the data, and the combination parameters found by the optimisation algorithm are only marginally optimal with respect to other parameter values. The locus (actually a valley) of marginal optimal parameters can be seen in the right hand side panels of \cref{fig: exact_SIR_fit}, where the cost function value of the optimisation algorithm is plotted as heat map. 
    
       \begin{figure}[H]
        \centering
        \includegraphics[width=1\textwidth]{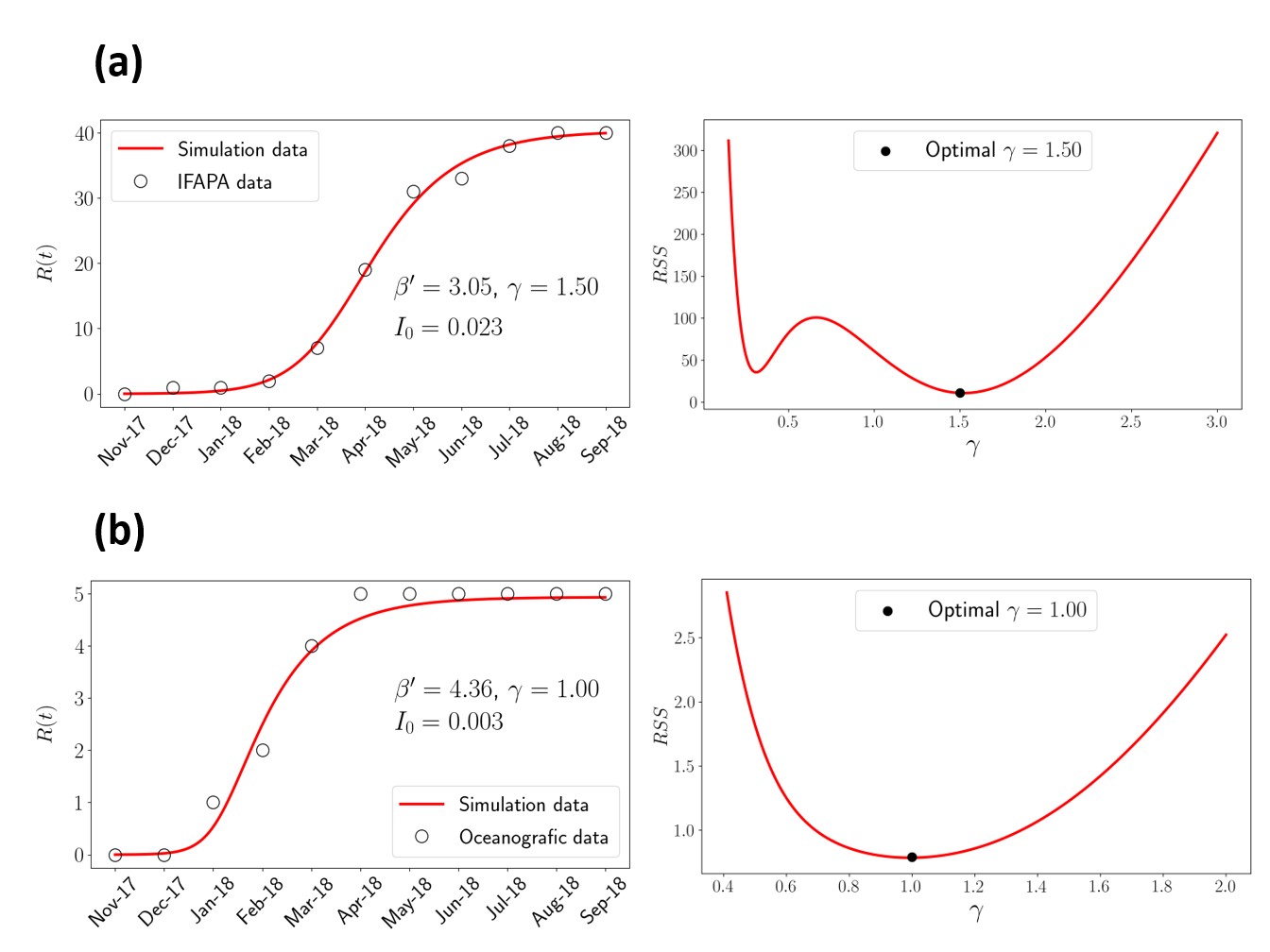}
        \caption{Parameter fitting for the R compartment to model (\cref{eq:SIRslow}) using data from IFAPA (panel (a)) and Oceanogràfic (panel (b)). The left part of both panels of the figure shows the optimal fit of the model to empirical data with $RSS=10.9, 0.8$ for IFAPA and OCE, respectively. The right panels show the variation of the $RSS$ error for some values of $\gamma$.  The $\beta'$ values have been obtained assuming a standard incidence, as explained in the main text.}
        \label{fig:approx_SIR_fit}
    \end{figure}
    
    Now we reach the point regarding the dilemma between mass action and standard incidence discussed in \cref{subsec:modtruct}. If one does not correct the $\bar{\beta}$ parameter with the size of the host population, $N$, that is equivalent to assuming the mass action incidence $\bar{\beta}=\beta$, the values that one would obtain for $\beta'=\beta\lambda/\mu=\lambda'/\mu$ for both populations take disparate values in both tanks: $\beta'=0.046$ for the IFAPA data set and $\beta'=0.87$ for the Oceanogràfic (OCE) data set, a factor of $19$ between them while their temperatures differ only by $3\,{}^o$C. 
    These numbers indicate that the standard incidence is more reasonable, what amounts to choosing $\bar{\beta}=\beta/N$, where the final values of the reported parameter $\beta'$ should be multiplied by $N=50$ for the IFAPA tank and $N=5$ for the OCE tank. The final result is then $\beta'=2.28$ and $\beta'=4.36$ for IFAPA and OCE tanks, that are the values reported in  \cref{fig: exact_SIR_fit}, implying that an almost twofold increase of the $\beta'$ parameter corresponds to an increase of $3\,{}^o$C. This relation is in good agreement with the typical changes in rates of a wide range of organisms with a $\SI{3}{\degree C}$ change in temperature, while a 19-fold change in the rate would imply at least a $\SI{30}{\degree C}$ change in temperature (cf. \ref{app:rate_change}).
    
    The fact that there is an infinite number of combinations of the parameters that optimally fit the real data suggests that, as two parameters are slaved one to each other, that the model
    admits a further reduction. This reduction corresponds exactly to the approximate $SIR$ model derived in \cref{eq:SIRslow}, with the relationship $\beta'=\lambda'/\mu$, as anticipated. So, this gives further corroboration to the use of the $SIR$ model \cref{eq:SIRslow} to fit $\beta'$ as the free parameter (fixing the value of $\gamma$ and with $I_0$ as the initial condition determined by the fit). For consistency with the previous fitting we expect to obtain $\beta'=2.28$ and $4.36$ as the optimal parameters for the IFAPA and OCE water tanks, respectively, and this is the case.
    
    Interestingly, as reduced model \cref{eq:SIRslow} has fewer parameters to fit we can relax our initial assumption of $\gamma=\SI{1}{month^{-1}}$ and check how the fit improves or worsens when varying $\gamma$.
    In \cref{fig:approx_SIR_fit} a fit of the reduced SIR model \cref{eq:SIRslow} is shown for the IFAPA (top) and Oceanogràfic (bottom) controlled water tanks\footnote{The $N$ correction corresponding to standard incidence has already been applied to these values.}. 
    \cref{fig:approx_SIR_fit}(c-d) shows the $RSS$ error as $\gamma$ is varied.
    It can be seen that for the IFAPA water tanks $\gamma=\SI{1.5}{month^{-1}}$ yields more accurate results, while for the Oceanogràfic water tanks $\gamma=\SI{1}{month^{-1}}$ remains optimum. This shows a decrease in the mean removal time $1/\gamma$ for lower water temperatures, with the finite size errors inherent to the OCE tank (as $N=5$).
    In the left panels the simulated curve of dead individuals, $R$ compartment, as a function of time for the optimal fitted parameters is confronted to the experimental data, showing a remarkable agreement. With the optimum values of $\gamma$, in the IFAPA tank (now with $\gamma=\SI{1.5}{month^{-1}}$) a new value of $\beta'=3.05$ is obtained, implying a probably more reasonable ratio of $1.43$ for $\beta'$ in both tanks (it was $1.91$ in the original fit).
    From the optimal parameters we obtain the basic reproduction number, since $R_0=\beta'/\gamma$ we have that $R_0^{\textrm{IFAPA}}\simeq2$ and $R_0^{\textrm{OCE}}\simeq4$, clearly above the epidemic threshold. 
    
    Summarising, the SIRP model is able to fit two sets of experimental data, agreeing with a standard incidence, according to which the infection rate depends on the amount of parasites per pen shell individual. \textit{Pinna nobilis} individuals in
    the IFAPA experiment were actually distributed in $4$ tanks, and the standard incidence is compatible with this experimental aspect. The temperature dependence of the fitted parameters in this range ($14-17 {}^o C$, appears to be compatible (although experiments at different temperatures would be needed) with an Arrhenius dependence of the infection parameters, also known as Boltzmann-Arrhenius 
    \citep{Brown2004,Molnar2017}, that can be extended to account for the expect unimodal dependence on temperature, with a maximum infectivity at a characteristic temperature for the parasite \citep{Molnar2017}. Therefore, we can assume that global change (or temperature shifts) is expected to have complex effects on infectious diseases, causing some to increase, others to decrease, and many to shift their distributions \citep{Rohr2020}. 
In the particular case of pen shell mortality, our model results suggest the proposed mechanism of lower disease expression at lower temperatures. This might have direct consequences for the development of the mortality event and offers a bleak perspective for the future and specifically in the eastern Mediterranean basin, where the mortality was observed later due to current patterns but average temperatures tend to be higher than in the western part of the Mediterranean. 

\section{Conclusions} \label{sec:conclusions}

In this work we have analysed a compartmental model to study marine epizootics for sessile hosts assuming infection by direct transmission through waterborne parasites. Moreover, we have used data from the recent mass mortality event of \textit{Pinna Nobilis} in the Mediterranean Sea as a case study to validate our model. Compartmental models are routinely used in the study of disease infection and propagation in terrestrial ecosystems, including the study of the current Covid-19 pandemic (see, e.g., \citep{Castro2020}). However, these models are starting to be used only recently in the study of marine epizootics \citep{article_SIP}, while proliferation models have been the most popular in the field \citep{Powell2015}. A reason for
the low popularity of compartment models in the study of marine epizootics is that there are some aspects in its modelling that differ from the now standard application to terrestrial ecosystems \citep{MCCALLUM_intro}. An important difference is that, in principle, (micro)parasites need to be modelled explicitly in marine ecosystems, while often they are not included in the description in terrestrial ecosystems  \citep{May1979}.

The SIRP model has $4$ compartments and depends on $4$ parameters, so that it is not quite amenable to theoretical analysis. At the same time, due to the large number of parameters of the model, using it to analyse experimental observations can be cumbersome in practice if the parameter values are unknown. Nevertheless, we have shown three reductions of the model, one exact and two approximate ones, that can be useful to overcome these limitations that are typically present at the first stages of emergent epidemics. Indeed, the timescale approximation is able to fit the collected data of our case study for some optimal parameters, as shown in \cref{sec:validation}. This approximation is particularly useful as it only depends on $2$ parameters, the death rate of infected hosts, $\gamma$ and an effective infection rate, $\beta'$. Although this approximation simplifies the fitting procedure, there is a price to be paid in this analysis. The infection parameter, $\beta$, and the parameters regulating proliferation, $\lambda$, and deactivation/dilution of the parasite, $\mu$, become entrained into a single effective parameter, $\beta'$. Thus, the full understanding of the different effects at play in the system requires further work. Furthermore, we have shown that an epidemic model for immobile hosts can be reduced to the standard SIR model, which assumes direct contact  among the hosts, i.e. that the hosts are mobile. This reduction is only valid when the time scale of the parasites is much faster than that of the hosts, i.e. $\mu\gg\beta N,\gamma$. Thus, our work provides a ground to apply the SIR model in marine epidemics of sessile hosts that fulfil the required conditions.

In a world with many possible new epizootics, we believe that our reduced model can be specifically useful to understand key features of those emerging diseases characterised by the spreading of waterborne parasites in a relatively fast way, provided that the temporal evolution of the disease can be determined for, at least, some set of individuals. Thus, some of the key parameters can be fitted to the available experimental data as shown in \cref{{sec:validation}}. Still, the fitted relevant parameters may need to be supplemented with further information or targeted experiments. We hope that this approach can be useful in understanding emerging diseases in shellfish species of economic not only ecological value, and also, with suitable modifications, in aquaculture. It is noteworthy that our case study is a haplosporidan waterborne parasite.  In fact, waterborne haplosporidans have been responsible for some of the most significant and consequential marine disease epizootics on record and are considered the major pathogens of concern for aquatic animal health shellfish industries around the world \cite{Arzul2015}. The SIRP model is the simplest model that one could think of having in mind its practical application, but could be extended to incorporate further effects that are so far described in an effective way. 

\section*{Acknowledgments}
We acknowledge Prof. Emilio Hern\'andez-Garc\'{\i}a and Eduardo Moralejo for useful comments and a critical reading of the manuscript. We acknowledge Dr. Maite V\'azquez-Luis for providing us with helpful information and remarks before setting up the SIRP model and Dr. José Ignacio Navas Triano for useful remarks about the interactions between bivalves and parasites. AGR and MAM acknowledge financial support from FEDER/Ministerio de Ciencia, Innovación y Universidades - Agencia Estatal de Investigación through the SuMaEco project (RTI2018-095441-B-C22) and the María de Maeztu Program for Units of Excellence in R\&D (No. MDM-2017-0711).

\appendix
\section{Finding a conserved quantity for the SIRP model} \label{app:P_exact}

   Starting with the SIRP model,
    \begin{equation}\label{eq:SIRP2}
        \begin{aligned}
            \dot{S} &=-\bar{\beta} P S  \\
            \dot{I} &=\bar{\beta} P S-\gamma I \\
            \dot{R} &=\gamma I \\
            \dot{P} &=\lambda I-\bar{\beta}P S -\mu P \ ,
        \end{aligned}
    \end{equation}    
    from the $\dot{S}$ equation, $P$ can be written as follows,
    \begin{equation}\label{eq:P_rel}
        P=-\frac{1}{\bar{\beta}}\frac{\dot{S}}{S} \ ,
    \end{equation}
    and summing up the equations for $\dot{S}$ and $\dot{I}$ the following relation for $I$ is obtained
    \begin{equation}\label{eq:I_rel}
        I=-(\dot{S}+\dot{I})/\gamma \ .
    \end{equation}
    Replacing \cref{eq:P_rel}, \cref{eq:I_rel} and the differential equation for $\dot{S}$ in the $4$th differential equation in \cref{eq:SIRP2} one obtains,
    \begin{equation}
    \dot{P}=-{{\lambda}\over{\gamma}} (\dot{S}+\dot{I})+\dot{S}+ {{\mu}\over{\bar{\beta}}}\ . {{\dot{S}}\over{S}}
    \end{equation}
    As $\dot{S}/S=d(\ln S)/dt$, all terms in the previous equation are exact differentials with respect to time, and the equation can be integrated yielding,
    \begin{equation}
    P + \frac{\lambda}{\gamma}\parentesi{S+I}-S-\frac{\mu}{\bar{\beta}}\ln S=C
    \label{eq:conservedquantity}
    \end{equation}
    with the integration constant $C$, 
    that is a conserved quantity, i.e., it takes the same value at one time of the dynamical evolution of the system.
    $C$ is related to the initial conditions by,
    \begin{equation}
        C = P(0)+\frac{\lambda}{\gamma}\parentesi{S(0)+I(0)} - \frac{\mu}{\bar{\beta}}\ln{S(0)} - S(0)=P(0)+\frac{\lambda}{\gamma}\parentesi{N-R(0)} - \frac{\mu}{\bar{\beta}}\ln{S(0)} - S(0)
        \label{eq:C_constant}
    \end{equation}
    
    It is possible to use \cref{eq:conservedquantity}-\cref{eq:C_constant} to express one of variables as a function of the others, for example 
   the parasite concentration $P$ as,
    \begin{equation} \label{eq:PSI_exact}
        P(S, I)=P(0)-\frac{\lambda}{\gamma}\left(S+I-N+R(0)\right)+ \frac{\mu}{\bar{\beta}}\ln{\frac{S}{S(0)}} + S - S(0)\ , 
    \end{equation}    
    or equivalently as,    
    \begin{equation}\label{eq:psrexact}
        P(S,R)=P(0) + \frac{\lambda}{\gamma}\claudator{R-R(0)+ \frac{\mu\gamma}{\bar{\beta}\lambda}\ln{\frac{S}{S(0)}}} + S - S(0)
    \end{equation}
    
       From \cref{eq:conservedquantity}, it is easy to show that the SIP model of Ref. \citep{article_SIP}, that differs from the SIRP model in that the
    fourth equation is simplified to $\dot{P}=\lambda I-\mu P$, has as 
    exact conserved quantity,
        \begin{equation}
    P + \frac{\lambda}{\gamma}\parentesi{S+I}-\frac{\mu}{\bar{\beta}}\ln S={\mathcal C}
    \label{eq:conservedquantitySIP}
    \end{equation}
    as the extra term in the SIRP model $-\bar{\beta}SP$ is equal to $\dot{S}$ from the first equation \cref{eq:SIRP2}.
    
    The SIR model has a conserved quantity \cite{Murray_book}, that in the case of
    \cref{eq:SIRslow} takes the form,
     \begin{equation}
I+S-{{\gamma}\over{\beta'}} \ln S =C\ .
\label{eq:SIRconservedquantity}
 \end{equation}
  Rewriting \cref{eq:conservedquantity} in the alternative form,
     \begin{equation}
    {{\gamma}\over{\lambda}} P +\left(1-{{\gamma}\over{\lambda}}\right) S + I -{{\mu\gamma}\over{\lambda\bar{\beta}}} \ln S=C'
    \label{eq:conservedquantity2}
    \end{equation}
 it can be seen that if $\lambda>>\gamma$ \cref{eq:conservedquantity2} reduces to
\cref{eq:SIRconservedquantity}, remembering that in \cref{eq:SIRslow}   $\beta'=\lambda\bar{\beta}/\mu$. The assumptions used to arrive to \cref{eq:SIRconservedquantity} in \cref{sec:fastslow} where $\mu>>(\gamma,\bar{\beta})$, and taking into account the expression for $R_0$ \cref{eq:R_0_SIRP}, that $\lambda\gtrsim\mu$ is most plausible to keep $R_0$ above the epidemic threshold ($R_0>1)$.

\section{Stability analysis of the fixed points of the SIRP model} \label{app:linstabfp}

Here we will assume the initial fixed point of our SIRP model, with $I(0)=P(0)=0$ right before the introduction of the infection, either through $I$ or $P$.
We will assume that $R(0)=0$, so that $S(0)=N$.
To study the linear stability of the model we need to write the Jacobian, that takes the form,
\begin{equation}
J=
\begin{pmatrix}
-\bar{\beta} P & 0 & 0 & \bar{\beta} S\\
\bar{\beta} P & -\gamma & 0 & \bar{\beta} S\\
0 & \gamma & 0 & 0 \\
-\bar{\beta} P & \lambda & 0 & (\bar{\beta} S-\mu)
\end{pmatrix}
\end{equation}
and obtain the eigenvalues for both fixed points, where we have already used the standard incidence, $\bar{\beta}=\beta/N$, from the evidence of the validation with experiments.
For the pre-epidemic fixed point, the Jacobian becomes,
\begin{equation}
\begin{pmatrix}
0 & 0 & 0 & \bar{\beta} S(0)\\
0 & -\gamma & 0 & \bar{\beta} S(0)\\
0 & \gamma & 0 & 0 \\
0 &\lambda & 0 & (\bar{\beta} S(0)-\mu)
\end{pmatrix}
\label{eq:Jacobian}
\end{equation}
Matrix \cref{eq:Jacobian} has two null $(0$) eigenvalues and a pair of eigenvalues given by,
\begin{equation}
\Lambda_{1,2}=-{1\over 2}\left(\gamma+\mu+{{\bar{\beta} S(0)}}\pm\sqrt{\gamma^2+\mu^2+\left({{\bar{\beta} S(0)}}\right)^2
+{{2\mu\bar{\beta} S(0)}}-2\gamma\mu
-{{2\gamma\bar{\beta} S(0)}}+{{4\lambda\bar{\beta} S(0)}}}
\right)
\end{equation}
from which one can determine that the fixed point is unstable whenever
\begin{equation}
\lambda\bar{\beta} S(0)>\gamma(\mu+\bar{\beta} S(0))
\label{eq:eigenvineq}
\end{equation}
and stable if the inequality is reversed.
It can be easily shown that \cref{eq:eigenvineq} is equivalent to $R_0>1$, with $R_0$ given by \cref{eq:R_0_SIRP}.

The final point of the epidemic,  $S(\infty)$, can be found by solving the transcendental equation,
     \begin{equation}
    \left({{\lambda}\over{\gamma}}-1\right)S(\infty)
   -\frac{\mu}{\bar{\beta}}\ln (S(\infty))=C
    \label{eq:sinftytreqn}
    \end{equation}
    where $C$ is determined from the initial conditions (\cref{eq:C_constant})
    and $I(\infty)=P(\infty)=0$.
    (\cref{eq:sinftytreqn}) has two roots, where $S({\infty})$ represents the smallest one.
 
 \section{Calculation of $R_0$ using the Next Generation Matrix method}\label{app:NGM}

The so called Next Generation Method (NGM) is a method to obtain $R_0$, the basic epidemiological quantity that measures the number of secondary cases produced by a typical infected individual during its entire period of infectiousness in a completely susceptible population. It was discussed in \ref{app:linstabfp} that $R_0$ is related to the largest non-zero eigenvalue, say $\Lambda$, of the fixed point corresponding to the infection-free equilibrium. An outbreak occurs when $\Lambda>0$ (or equivalently when $R_0>1$) and the NGM is an ingenious method to obtain directly $R_0$ in a reduced linear system.
In more concrete terms, within the NGM method $R_0$ is the dominant eigenvalue of a suitably defined linear operator (a linear matrix in a suitable basis). This operator is obtained from a decomposition of the Jacobian, $\bold J$ of the infected/infecting compartments (i.e. excluding susceptible and removed compartments) in the form $\bold J=\bold T+\bold\Sigma$, where $\bold T$ is the \textit{transmission part}, that describes the production of new infections, and $\bold \Sigma$  the \textit{transition part}, that describes changes of state (including death). Then, it can be proved \citep{Diekmann2010} that the \textit{basic reproduction number} $R_0$ is given by the spectral radius (i.e. the largest eigenvalue) of the (next generation) matrix $\bold K=- \bold T \bold \Sigma^{-1}$.

In the case of the SIRP model the decomposition is applied to the $2\times 2$ Jacobian corresponding to the dynamical evolution of the $(I,P)$ \textit{infectious} compartments, being the decomposition,
\begin{equation*}
    \bold J=\parentesi{\begin{array}{cc}
    -\gamma & \bar{\beta} S_0 \\
    \lambda & -(\bar{\beta} S_0+\mu) \end{array}}\qquad
    \bold T=\parentesi{\begin{array}{cc}
        0 & \bar{\beta} S_0 \\
        0 & 0
    \end{array}} \qquad \bold\Sigma=\parentesi{\begin{array}{cc}
        -\gamma & 0 \\
        \lambda & -(\bar{\beta} S_0 + \mu)
    \end{array}}
\end{equation*}
where the $\bar{\beta} PS$ term in $\dot{I}$ is the only one that contributes to the transmission matrix, as it is the only process involving infection, while all the other terms in the dynamical equations of $\dot{I}$ and $\dot{P}$ imply transitions (to another compartment, like $I\rightarrow R$ or birth and death of $P$).

Then, the next generation matrix is given by,
\begin{equation*}
    \bold K=-\bold T\bold\Sigma^{-1}=\parentesi{\begin{array}{cc}
          \displaystyle \frac{\lambda\beta S_0}{\gamma(\beta S_0 + \mu)} & \displaystyle \frac{\beta S_0}{\beta S_0 + \mu} \\
        0 & 0
    \end{array}} \Longrightarrow R_0=\frac{\lambda \beta S_0}{\gamma\parentesi{\beta S_0+\mu}} \ ,
\end{equation*}
This result coincides with the expectation that $R_0$ should correspond to the number of  hosts infected in a single generation by the appearance of an infected host in a completely susceptible population. This can be obtained from the number of parasites produced by an infected host, $\lambda$, times the time in which the infected host is alive producing parasites, $1/\gamma$, multiplied by the number of infected hosts produced per parasite, $\beta S_0$, times the time the parasite is alive available to infect, $1/(\mu+\beta S_0)$, taking into account that parasites are inactivated at a rate $\mu$ and also die when infecting at a rate $\beta S_0$, where this result assumes that the susceptible population does not change from its initial value $S_0$.

\section{Sensitivity Analysis} \label{app:sensanal}

 One particular way to analyse the local sensitivity (LSA) of a given model function, $F(\vec{p})$, for each of the parameters that conform it, $p_i$, is through the normalised sensitivity indexes \citep{sensitivity_analysis}, 
    \begin{equation}\label{eq:sensitivity_analysis}
        \Omega_{p_i}^{F}=\dpart{F}{p_i}\frac{p_i}{F}\Big|\limitss{p_i=p^0}{} \ .
    \end{equation}
    where the partial derivatives in \cref{eq:sensitivity_analysis} are determined analytically in our case.
    
    GSA works by studying the influence of a large domain of parameter space in the final state of the epidemic and in the epidemic peak.
   In our case this will be achieved by means of a variance based analysis, known as Sobol method \citep{SOBOL2001271}. This particular method provides information no only on how a particular parameter alone influences the model outputs (as happens with LSA), but also on the influence of its interactions with other parameters. This information is organised in what are known as Sobol indices, that have been
   implemented within the Julia high-level programming language \citep{julia} using the DifferentialEquations.jl package \citep{DifferentialEquations.jl}, and in particular through its subpackage DiffEqSensitivity.jl. This implementation allows the user to sample the parameter space using QuasiMonteCarlo methods and thus obtain confidence intervals (CI) for the sensitivity indices, which are directly related to the committed statistical error.
   
   The total order indices are a measure of the total variance of the output quantity caused by variations of the input parameter and its interactions. First order (or ``main effect'') indices are a measure of the contribution to the output variance given by the variation of the parameter alone, but averaged over variations in other input parameters. Second order indices take into account first order interactions between parameters. Further indices can be obtained, describing the influence of higher-order interactions between parameters, but these are not going to be considered.
    More detailed information about sensitivity analysis can be found in \citep{Sensitivity_analysis_book}.

\section{General rate change with temperature} \label{app:rate_change}

    In \citep{Gillooly2248} the metabolic rate of a wide variety of organisms was studied, showing that the change in the metabolic rate with temperature was similar among them. In particular, the natural logarithm of the metabolic rate linearly depends on the inverse of absolute temperature,
    \begin{equation}
        \log(R(T))=a\cdot\parentesi{\frac{100}{T}}+b
    \end{equation}
    and for all the analysed organisms they found that $a$ lies between $-5$ and $-10$ and $b$ between $14$ and $30$. From their analysis, we can compute the change in the rate for a given increase of temperature,
    \begin{equation}\label{eq:rate_change}
        \frac{R(T+\Delta T)}{R(T)}=\exp(a\cdot\parentesi{\frac{100}{T + \Delta T}}+b) / \exp(a\cdot\parentesi{\frac{100}{T}}+b) = \exp(a \cdot \frac{-1000}{T+\Delta T}\cdot \frac{\Delta T}{T}) \ .
    \end{equation}
    Substituting $T=287 K$ and $\Delta T=3 K$, that correspond to our available data (cf. \cref{sec:validation}) in \cref{eq:rate_change}, using both the upper and lower limit of $a$, we obtain that the expected increase in the effective transmission rate is between $1.2$ to $1.4$. This is far from the $19$-fold increase that we obtained with the mass action hypothesis in \cref{sec:validation}
    while it is in good agreement with either the $1.92$ ratio we obtained  for $\bar{\beta}$ with the reduction of \cref{sec:exactredapp} or the $1.43$ ratio obtained with the fast-slow approximation of \cref{sec:fastslow}, both obtained using the standard incidence choice. 
    
\begin{figure}[H]
        \centering
        \includegraphics[width=1\textwidth]{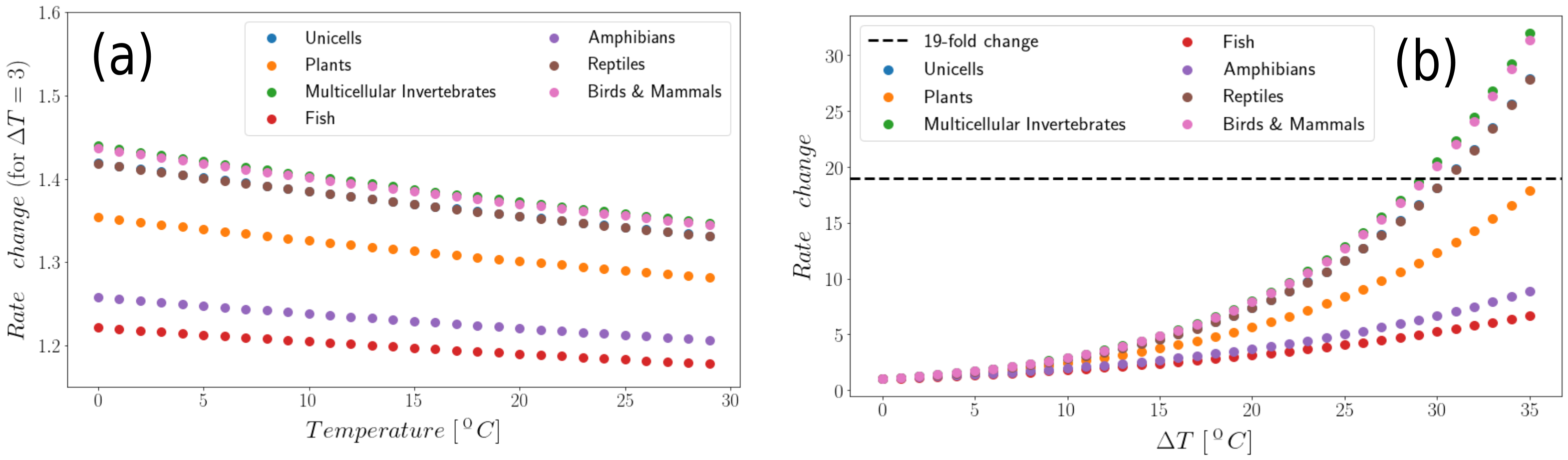}
        \caption{Graphical representation of change in the rate (in ordinates) for different reference temperatures (in abscissae) for: (a) a temperature increase of $\SI{3}{\degree C}$; (b)  a temperature increase of $\SI{14}{\degree C}$. The black dotted line in (b) corresponds to a $19$-fold increase in the rate.
        }
        \label{fig:rate_changes}
    \end{figure}

    \cref{fig:rate_changes}(a) shows the change in the rate with an increase of $\SI{3}{\degree C}$ for different base temperatures and for all the organisms analysed in \citep{Gillooly2248}, and using their fit. Note that for all temperatures between $\SI{0}{\degree C}$  and $\SI{30}{\degree C}$ the rate change lies between $1.2$ and $1.45$. \cref{fig:rate_changes}(b) shows the change in the rate for different temperature increases, with a base temperature of $T=\SI{287}{K}$. Note that in order to obtain a $19$-fold increase the temperature change should be at least of $\SI{30}{\degree C}$\footnote{A temperature change of $\SI{30}{\degree C}$ could fall outside the range in which the study of \citep{Gillooly2248} is valid. We just stress that a $19$-fold rate change is unlikely for the case of a $\SI{3}{\degree C}$ that correspond to the $2$ data sets that we compare in this section.}.
   The temperature dependence of metabolic rates has been reported in the context of epidemic parameters \citep{COELHO2006,Shapiro2017}
    
    The behavior of the metabolic rates re-analysed here has been also found experimentally in epidemic contexts such as \citep{COELHO2006, Shapiro2017}, i.e. the increase of the rates with temperature fulfill the ranges shown here.
    
\newpage

\end{document}